\documentclass[12pt]{article}

\usepackage[T1]{fontenc}
\usepackage[utf8]{inputenc}
\usepackage[left=3cm,right=3cm,top=2cm,bottom=2cm]{geometry}

\usepackage{latexsym}
\usepackage{amssymb,amsbsy,amsmath,amsfonts,amssymb,amscd,mathrsfs,dsfont}
\usepackage{graphicx,color}
\usepackage{float}
\usepackage{caption}
\usepackage{subcaption}
\usepackage{hyperref}
\usepackage{cancel}
\usepackage{tikz}
\usetikzlibrary{shapes,shadows,arrows}
\usepackage[toc,page]{appendix}
\usepackage[T1]{fontenc}
\usepackage[style=numeric,    maxbibnames=15, bibencoding=utf8, backend=bibtex, sorting=none]{biblatex}
\DefineBibliographyStrings{english}{%
    andothers = {\em et\addabbrvspace al\adddot}
}
\bibliography{BiblioMC_09_03}

\newcommand{\posA}{x}
\newcommand{\tempsA}{t}

\newcommand{\FluxForwardGenerique}{\Phi_{\!f}}
\newcommand{\FluxBackwardGenerique}{\Phi_b}
\newcommand{\CoeffDiffGenerique}{c}
\newcommand{\CoeffTranstFToBGenerique}{r}
\newcommand{\FixedGenerique}{\Psi}
\newcommand{\CoeffUtilisation}{u}
\newcommand{\FuncFixGenerique}{F}
\newcommand{\CoeffFixGenerique}{f}
\newcommand{\UsedGenerique}{\Xi}
\newcommand{\SortieGenerique}{O}
\newcommand{\DomaineIntSortieGenerique}{\Omega}
\newcommand{\SourceExterieure}{Q}
\newcommand{\Velocity}{\omega}
\newcommand{\EFfracp}[2]{\frac{\displaystyle\partial#1}{\partial#2}}
  
\newcommand{\VelocityOpt}{\omega_{{Opt}}}
\newcommand{\CoeffTranstFToBGeneriqueOpt}{r_{{Opt}}}
\newcommand{\CoeffFixGeneriqueOpt}{f_{{Opt}}}
\newcommand{\CoeffUtilisationOpt}{u_{{Opt}}}
  
\newcommand{\Seuil}{L}
\newcommand{\Bsup}{Upp}
\newcommand{\Binf}{Low}
  
\newcommand{\fracp}[2]{\frac{\partial #1}{\partial #2}}

\newcommand{\WorthOneInZeroOnBorder}{\chi}

    \makeatletter
    \def\ps@pprintTitle{%
      \let\@oddhead\@empty
      \let\@evenhead\@empty
      \let\@oddfoot\@empty
      \let\@evenfoot\@oddfoot
    }
    \makeatother

\begin{document}



\begin{center}
\Large{An innovative Statistical Learning Tool based on Partial Differential Equations for livestock Data Assimilation}\\[0.7cm]

\normalsize{}
H\'{e}l\`{e}ne Flourent$^{1,2,}\footnote[4]{\href{mailto:helene.flourent@univ-ubs.fr}{helene.flourent@univ-ubs.fr}}
$, Emmanuel Fr\'{e}nod$^{2,3,}\footnote[5]{\href{mailto:emmanuel.frenod@univ-ubs.fr}{emmanuel.frenod@univ-ubs.fr}}
$, Vincent Sincholle$^1$\\[0.7cm]

\textit{~$^1$ NutriX\footnote[6]{The compagny wishes to remain anonymous}, France}

\textit{~$^2$ Universit\'{e} Bretagne Sud, Laboratoire de Math\'{e}matiques de Bretagne Atlantique,UMR CNRS $6205$, Campus de Tohannic, Vannes, France}

\textit{~$^3$ See-d, $6$, rue Henri  Becquerel - CP $101$, $56038$ Vannes Cedex, France}

\end{center}

\begin{abstract}
Realistic modeling of biological mechanisms requires a large volume of prior knowledge and leads to heavy mathematical models. On the other hand, the classical Machine Learning algorithms, such as Neural Networks, need a large quantity of data to be fitted. Nevertheless, to predict the evolution of biological variables we are often facing a lack of knowledge and a lack of data, especially in the livestock sector. Therefore, we explored an intermediate approach, called "Data-Model Coupling". We demonstrated that parametrized Partial Differential Equations (PDEs) can be embedded in a data fitting process and then in an efficient predictive Statistical Learning tool. We postulated that all the physico-chemical phenomena occurring in an animal body can be summarized by the circulation, the evolution and the action of an overall information flow. We built the PDE system which mathematically translates our assumption and we fitted it on data.\par
The applications of our approach to data relative to the growth of farm animals showed that it increases the forecasting accuracy and reduces the training data dependency of the resulting predictive tool. Moreover, learning the dynamics linking the inputs and the outputs confers to the tool the capability to be trained on a given range of data and then to be accurately applied outside this range of data. This extrapolation capability is a real improvement over existing predictive tools.
\\

\noindent\textit{keywords} : Statistical Learning, PDE, Forecasting,  Data Assimilation, Data-Model Coupling, Biological Mathematical Modeling.

\end{abstract}


\section{Introduction}
\label{S:1}

Smart Farming corresponds to the use of new technologies to make the farm production processes more efficient.\par
As it can be identified in \cite{McPhee2009MathematicalMI}, \cite{puillethal01137029}, \cite{martin_sauvant_2010}, \cite{Nkrumah2007}, \cite{Nesetrilova2005} and \cite{Basarab2003}, in the agri-food sector, simulating and predicting the effects of nutrition on animal performances are two decisive and strategic goals for breeders and companies to optimize animal efficiency. However, the biological phenomena linking the nutrition and the performances of animals are complex. Furthermore, in most cases, to build tools able to predict the evolution of biological variables, it is necessary to jointly manage the complexity of the phenomena occurring in the studied biological system and the lack of data available to fit those tools. \par

Data Assimilation is an approach that embeds mathematical theories, Data Science and Computer Science processes to estimate the most likely state of a connected system at an instant $t$ (See \cite{auroux2005back}, \cite{gregg2009skill}, \cite{lguensat2017analog} and \cite{lguensat2019data}). To do so, it combines the information given by a predictive tool and the one contained in a more or less continuous stream of collected data. To very briefly sum up, it consists of considering that data flows are gathered to correct at a given frequency the simulation done by the predictive tool. This correction takes into account that collected data contain noise and the predictive tool embeds an intrinsic model error.\par
This combination of information could permit to know the state of an animal or a group of animals, in terms of health and performances, according to their ingestions and the drugs that are administered to them. Hence, this concept constitutes an interesting and promising way to oversee future livestock and address the Smart Farming issues (\cite{LI20111595}, \cite{rijk2013integration} and \cite{janssen2015towards}).\par
Biological data are not easy to collect and generally contain a large variability (\cite{Locke2005ModellingGN} and \cite{Yanjun2006}). Hence, to perform Data Assimilation in the livestock sector it is necessary to develop efficient and light predictive tools able to be fitted on few and scattered data relative to complex phenomena.\\

According to V{\'a}zquez-Cruz et al. \cite{VazquezCruz2014}, there are currently two general approaches to build tools predicting biological responses.\par
On the one hand, realistic modeling of biological mechanisms requires a large volume of prior knowledge and generally leads to heavy mathematical models (\cite{Bastianelli2} and \cite{Martin}). However, the complex implementation of these models limits their adaptability, in particular when it comes to processing or assimilating field data.\par
On the other hand, the structure of classical Machine Learning (ML) algorithms, such as Neural Networks, have limited ability to take into account the existence of complex underlying phenomena and need to be fitted on a large quantity of data to compensate for the absence of prior biological expertise (\cite{tan2003empirical}, \cite{Shavlik1995}, \cite{Hubbardnetworks} and \cite{dumpala2017kffnn}).\par
Hence, due to their lack of adaptability or their inability to be fitted to few data the existing tools are not entirely appropriate for achieving Data Assimilation in the context of "Biological Small Data".\\

We assumed that a global and synthetic consideration of the biological processes may help gain precision, in comparison with a classical ML tool which integrates no prior knowledge. We also assumed that this synthetic consideration permits us to do it while keeping a flexible and light tool, in comparison with a tool based on realistic models. Therefore, we explored an intermediate approach, named "Data-Model Coupling" to build predictive tools able to deal with both the complexity of the biological responses and the current lack of data.\par
This emerging approach is midway between the realistic modeling and the "Black Box" approach. As seen in \cite{frenod:hal-00817522}, \cite{rousseau2013modelisation}, \cite{sacks_coupling_2007} and \cite{wang2010physiological}, Data-Model Coupling approach consists of building a mathematical model, corresponding to a mathematical synthesis of the studied system. Then, the parameters contained in the model are fitted to data. As in the above-cited studies, the construction of our tool is based on an optimal combination of knowledge, to design a relevant mathematical model, and data, to optimize the model parameters. In our approach, the mathematical model is a wisely designed parametrized PDE system.\\

Data Assimilation is a long-term objective of our research work. Nevertheless, to finally obtain a tool particularly suitable to perform Data Assimilation, this long-term objective has strongly guided the whole modeling approach introduced in this paper.\\

Several contributions can be identified in this paper.\par
\begin{figure}[h]
    \centering
        \includegraphics[width=0.5\textwidth]{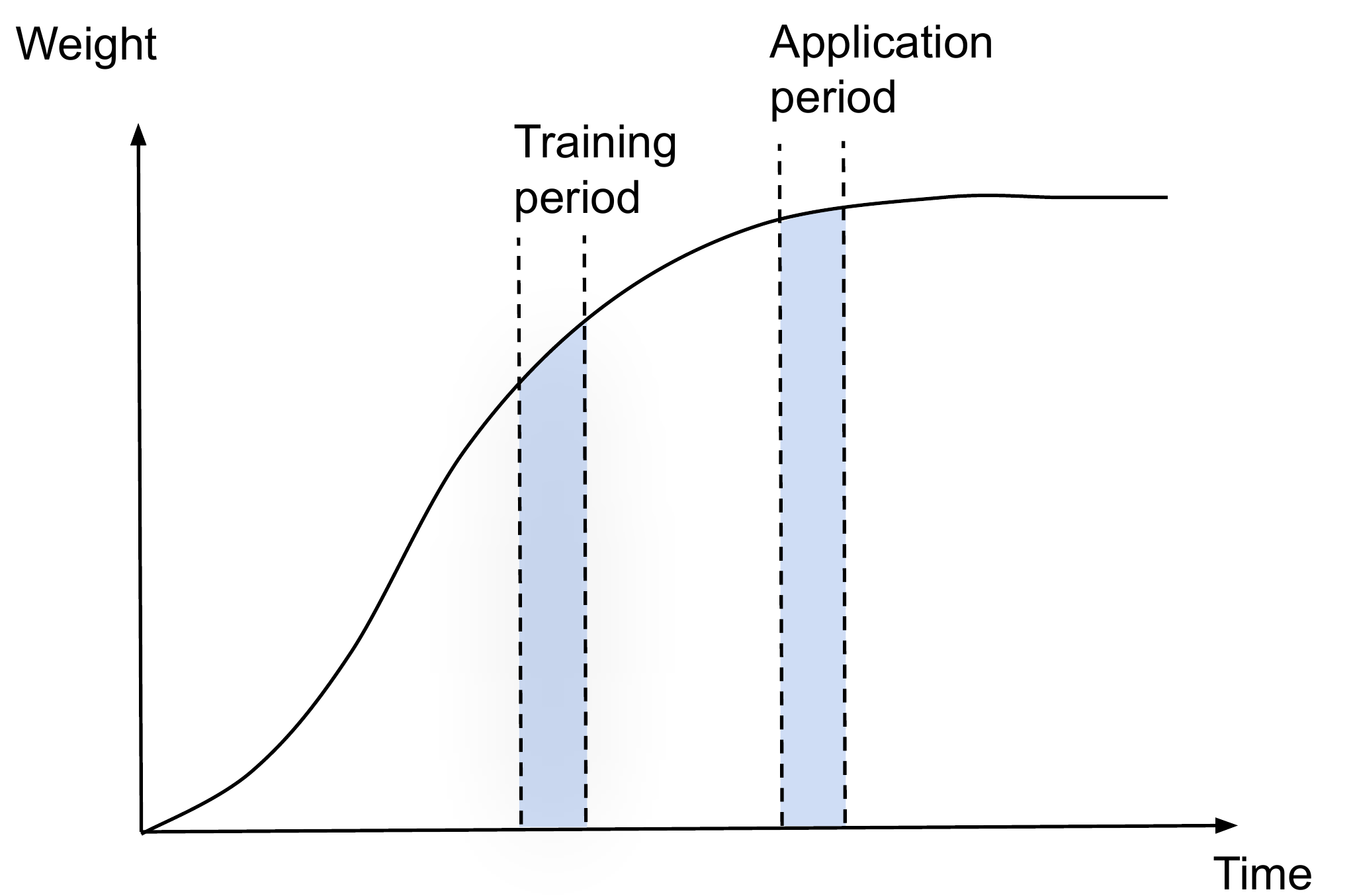}
    \caption{Schematization of the extrapolation capability of the developed tool}
    \label{fig:schema_extrapol}
\end{figure}

In the first place, the applications of this tool, on collected data relative to the growth of farm animals, put in evidence its extrapolation capability which is a real improvement over existing predictive tools. As it is illustrated in Figure \ref{fig:schema_extrapol}, in the application presented in Section \ref{Sec:Appli}, our tool was trained on a short training period to link the inputs and the outputs. Yet, it can accurately predict the outputs from the inputs outside the range of the Training Data. This is the peculiarity of our approach: the model learns synthetic dynamics linking the inputs and the outputs by fitting parameter-dependent evolution equations. Once the parameters fitted, those dynamics can be applied outside and even far from the Training Data range.\par
This extrapolation capability permits to reduce the amount of data to collect and thus to reduce the costs relative to the experiments and the data management. It also permits to extend the validity period of the prediction provided by the tool. Hence, when it will come to Data Assimilation issues, in perspective with what explained above, the correction of the prediction via the use of data could be less frequent and thus the computational costs could be lower.

\begin{figure}[h]
    \centering
        \includegraphics[width=1\textwidth]{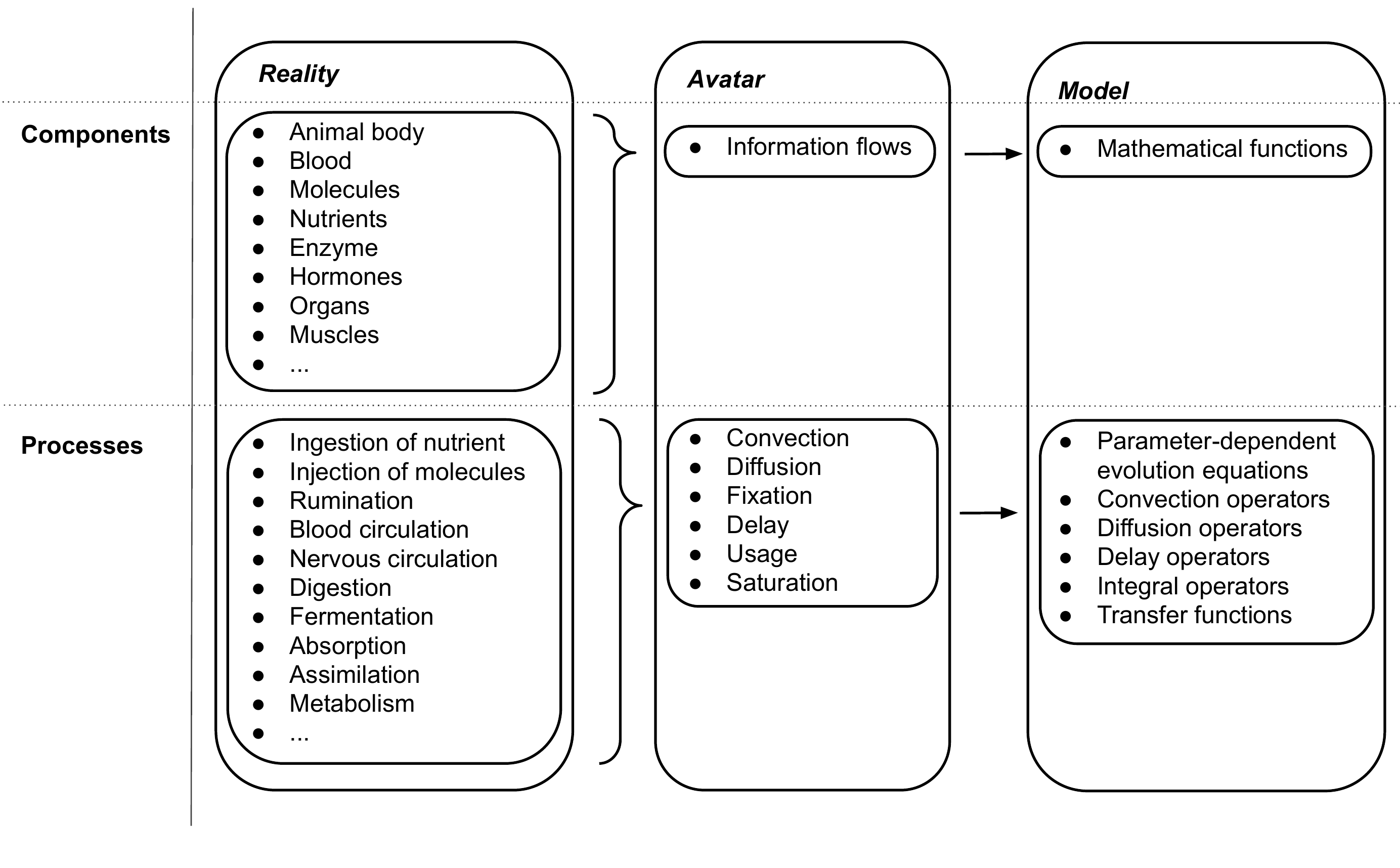}
    \caption{Schematization of the different sets considered in our approach and their own components and processes.}
    \label{fig:schema_real_mod}
\end{figure}

The second contribution of our exploration is the development of a concept between the reality and the model (Figure \ref{fig:schema_real_mod}). In most cases, the objective of a mathematical model is to translate the reality, adopting a higher or lower abstraction level. In our approach, a differentiation between the reality and our model is made. Indeed, the used support of reflection is not directly the real animal, but an Avatar which conceptually and essentially outlines the global dynamics occurring in the animal body. A large number of physico-chemical phenomena occur in the animal body in response to the ingestion or the injection of molecules. They lead, some time later, to the change of biological variables. This supply of molecules and those biological variables can be monitored and recorded to generate Input and Output data. We assumed that this kind of Inputs and Outputs can be linked by a dynamical model which is a mathematical translation of the Avatar. Therefore, we designed the PDE system mathematically translating our assumption and describing the convection, the diffusion and the action of an overall information flow.\par
Thirdly, the application of our approach on real data showed that our tool can accurately link biological-related Inputs and Outputs, even if it is fitted on few, scattered and noisy data.\par
Data-Model Coupling is so far essentially used in the fields of meteorology (see  \cite{Simmons_Hollingsworth_2002}), hydrology  (see  \cite{KIM2002393},\cite{Crosson_Laymon_2002} and \cite{MACKAY2003230}), biogeochemistry (see  \cite{Barrett_Damian_2002}, \cite{Barrett_Hill_2005}, \cite{Rayner_Scholze_2005}  and  \cite{SACKS_SCHIMEL_2006}) and oceanography (see \cite{ailliot:hal-00129093}). The successful use of a Data-Model Coupling approach to treat biological issues can be also considered as a contribution of this paper.\\

In this paper, we will show that the use of a short and relevant PDE system in a fitting process leads to the construction of an efficient predictive tool having a low data dependency and a high information extraction capability.\par
This tool can be used to predict the evolution of biological variables according to the ingestions and the injections of molecules in the animal body. The objective of the application presented in this paper was to predict the growth of two groups of animals of a specific species, according to their initial weight and their feeding behavior. But the genericity and the parsimony of our tool might ensure its suitability to predict other performance indicators relative to other farm species. \par
This low data dependency and this high information extraction capability allow the use of few data to fit our tool. Therefore it can be used to reduce the costs relative to experiments, data collection, and data storage. Furthermore, in comparison with the existing predictive tools, these capabilities also make our tool more suitable to efficiently perform Data Assimilation, even if the frequency of data collection and the quality of the collected data are low.\\

\begin{figure}[h]
    \centering
        \includegraphics[width=0.8\textwidth]{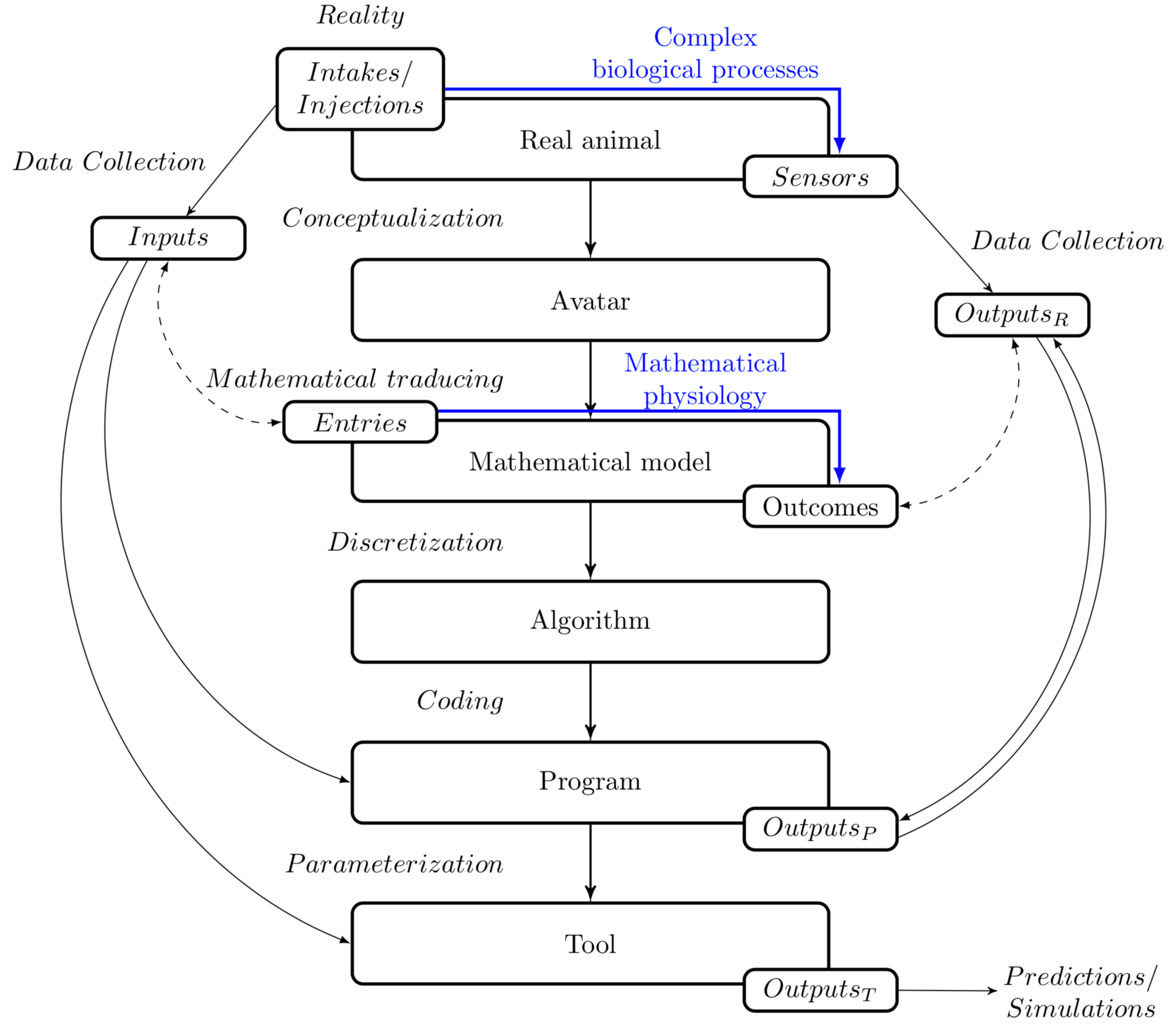}
    \caption{Articulation of the different elements of ourcapabilities exploration}
    \label{fig:Articulation}
\end{figure}

To summarize, in our approach we distinguished different dimensions. As it is illustrated by Figure \ref{fig:Articulation}, there is the \textit{Reality} in which there are \textit{Intakes} and \textit{Injections} inducing complex biological processes in the animal body. Some \textit{Sensors} extract information from this \textit{Reality} which is stored in databases made of \textit{Inputs} and \textit{Outputs}. Since the model is not directly assimilable to the reality, the inflows and the outflows of the model are also not directly assimilable to the input and the output data. The \textit{Inputs} have to be translated by a mathematical function into \textit{Entries}, that are pieces of information integrated into the \textit{Mathematical Model} and that induce the generation of \textit{Outcomes}, also linked to the \textit{Outputs} extracted from the \textit{Reality} by a mathematical function.

As it can be noticed in Figure \ref{fig:Articulation}, our exploration relies on the relationship between several diverse elements such as the \textit{Real Animal}, the \textit{Avatar} and the \textit{Mathematical Model}. The \textit{Algorithm} comes out of the discretization of the \textit{Mathematical Model}, e.g. the PDE system which mathematically translates what takes place in the \textit{Avatar}. This system of PDEs contains parameters corresponding to biological-like factors that can be learned from a database. The $Program$ corresponds to the code that manages this learning step. It uses an iterative training process during which an optimization algorithm finds the values of the parameters that minimize the difference between the measured and the predicted \textit{Outputs}. The $Tool$ finally corresponds to the \textit{Mathematical Model} parameterized with the values of the parameters obtained at the end of the learning step.\par
The presence of parameters that can be learned from data in the mathematical model confers learning ability to the tool based on this model. Using a database, we obtained a tool able to reconstitute dynamics between inputs and outputs to perform forecasts and extrapolations. Hence, the constructed tool can be considered as a \textit{Statistical Learning Tool}.\\

In this paper, we will present our modeling approach, the conception of our tool and the results of the applications of our approach on fictitious and real data.\par
After this Introduction, putting this research work in its proper context, we will detail in Section $2$ the conception of the mathematical model and its applicability.\par
We tested the well-functioning and the capacities of our tool in two different ways. First, we established a fitting method taking into account the relations existing between the model parameters and we generated a database to test this fitting method on it. This first application on mastered data allowed us to verify the ability of our tool to fit the parameters. Those simulation tests are presented in Section $3$. After those tests on fictitious data, we applied our approach to data collected on a farm and relative to the feeding behavior and the growth of two groups of animals. The results are presented in Section $4$. This application demonstrated the prediction capability of the tool in real conditions.\par
We put in evidence the potential of our tool and the improvement conferred by it. To do so, we compared the capabilities of our tool with the ones of some Logistic Models, Mechanistic Models and Machine Learning algorithms. These comparisons will be detailed in Section $5$.

\section{Construction and description of the {Mathematical Model}}

In our approach, particular attention was paid to the construction of the \textsl{Mathematical Model} embedded in the final predictive tool. Indeed, the designing of this model - that is a PDE system - was the key element to achieve our objectives of lightness, accuracy and learning potency.\par

\subsection{Conception of the {Mathematical Model}}
\label{SubSect:DescriptionModel}

Through the conceptualization of the \textit{Avatar}, we set up a parsimonious summary of any biological process. Indeed, we mathematically summarized the global intern dynamics of the animal via several equations and mathematical operators which we assumed necessary and sufficient.

\begin{figure}[h]
    \centering
    \includegraphics[width=0.75\textwidth]{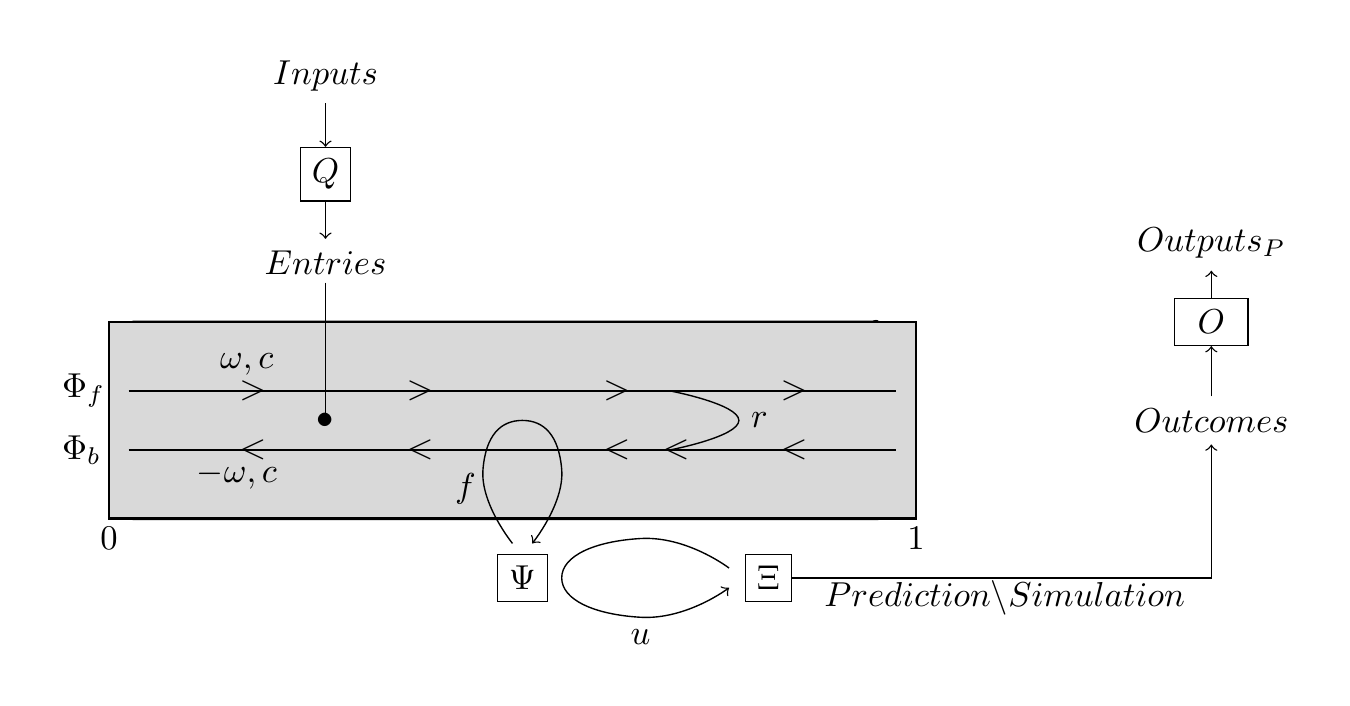}
    \caption{Schematization of the {Mathematical Model}}
    \label{fig:Schematization}
\end{figure}

We hypothesized that, when a molecule or a group of molecules enter the body of a living organism, it circulates in the body through a network of vessels containing a fluid. It integrates this fluid and uses it as a vector to evolve via convection and diffusion mechanisms. In the network of vessels, the molecules may be in competition with other mechanisms which may delay its progression. The circulating molecules may then be captured and accumulated in an organ or a specific tissue. During its storage, the molecules can be used and induce a change in some biological variables. Then, we built the PDE system which mathematically translates the previously set up summary illustrated by Figure \ref{fig:Schematization}.\\

We modeled our \textit{Avatar} using variables, densities, and fields that are all unitless and dimensionless. We also reduced the considered geometrical space to interval $\left[0~;1\right]$. We considered a Forward Flow $\Phi_f$, and a Backward Flow $\Phi_b$ streaming in this one-dimensional geometrical space. As it was described in the introduction, these flows can be seen as a very synthetic summary of blood circulation, a circulation in the nervous system or circulation in the digestive tract, according to the problematic.\par The involved {Inputs} in our tool can correspond to collected data relative to feed intakes, water intakes, and administered drugs. These {Inputs} are integrated in the {Mathematical Model} via a function $Q$ transforming these {Inputs} into information inflows, called \textit{Entries}. We modeled that part of the injected information circulates in a forward direction, via $\Phi_f$ and the rest circulates backward, via $\Phi_b$. This information can evolve via convection and diffusion phenomena. We assumed that the circulating information can be delayed, captured, stored and used to ultimately induce a modification in the \textit{Outcomes} $O$. These \textit{Outcomes} correspond to the model outflows. Those Outcomes are transformed by a mathematical function to be comparable with collected outputs.\\ 
\noindent Therefore, $\FluxForwardGenerique \! (\tempsA, \posA)$ and $\FluxBackwardGenerique \! (\tempsA, \posA)$ are, at each instant $t$, two space densities respectively associated with a forward flux with a velocity $\Velocity$ and a backward flux with a velocity $-\Velocity$.

\noindent The spatial density $\FluxForwardGenerique \! (\tempsA, \posA)$ is supposed to be solution to: 
        \begin{multline}
            \label{201608221651}
\EFfracp{\FluxForwardGenerique}{\tempsA} (\tempsA, \posA) + \Velocity \EFfracp{\FluxForwardGenerique}{\posA} (\tempsA, \posA)
-      \CoeffDiffGenerique\EFfracp{\bigg[
\WorthOneInZeroOnBorder \EFfracp{
        \Big[  \FluxForwardGenerique 
        + \FluxBackwardGenerique  \Big]}{\posA}
    \bigg]}{\posA} (\tempsA, \posA)~
\\
= \frac12 \SourceExterieure(\tempsA, \posA)
- \CoeffFixGenerique \FuncFixGenerique  (\posA) \FluxForwardGenerique  (\tempsA, \posA)
-  \CoeffTranstFToBGenerique \FluxForwardGenerique  (\tempsA, \posA),
        \end{multline}
        
\noindent Similarly, $\FluxBackwardGenerique(\tempsA, \posA)$ is supposed to be solution to:
        \begin{multline}
        \label{201608221652}
\fracp{\FluxBackwardGenerique}{\tempsA} (\tempsA, \posA) - \Velocity \EFfracp{\FluxBackwardGenerique}{\posA} (\tempsA, \posA)
-  \CoeffDiffGenerique\EFfracp{\bigg[
    \WorthOneInZeroOnBorder \EFfracp{
        \Big[  \FluxForwardGenerique 
        + \FluxBackwardGenerique  \Big]}{\posA}
    \bigg]}{\posA} (\tempsA, \posA) ~
\\
= \frac12 \SourceExterieure(\tempsA, \posA)
- \CoeffFixGenerique \FuncFixGenerique  (\posA) \FluxBackwardGenerique  (\tempsA, \posA)
+  \CoeffTranstFToBGenerique \FluxForwardGenerique  (\tempsA, \posA),
        \end{multline}
In these equations, the parameter $\CoeffDiffGenerique$ is the diffusion velocity of the information. The space-time density $\SourceExterieure$, corresponds to an external source of information. The function $\FuncFixGenerique$ is worth $0$ in certain areas of the involved geometrical space and $1$ in others. The area where this function is worth $1$ corresponds to the location of the entity capturing the information. The parameter $\CoeffFixGenerique$ determines the rate of fixed information. The parameter $\CoeffTranstFToBGenerique$  determines the fraction of the circulating information transferred from the Forward Flow to the Backward Flow, which induces a delay in the progression of the information.\\
The function $\WorthOneInZeroOnBorder$ is compactly supported in $(0,1)$, mainly constant and worthing $1$. This function integrated into the diffusion term makes diffusion vanish at the edges of the domain. At each time $t$, the spatial density $\FixedGenerique  (\tempsA, \posA)$, associated with the fixed information, is solution to:
\begin{gather}
            \label{201608221653}
            \fracp{\FixedGenerique}{\tempsA} (\tempsA, \posA) 
            = \CoeffFixGenerique  \FuncFixGenerique  (\posA)  \bigg[\FluxBackwardGenerique  (\tempsA, \posA) + \FluxForwardGenerique  (\tempsA, \posA)\bigg]
            - \CoeffUtilisation \FixedGenerique  (\tempsA, \posA).
\end{gather}
\noindent The parameter $\CoeffUtilisation$ is the coefficient determining the usage rate of the fixed information. At each time $t$, the spatial density $\UsedGenerique  (\tempsA, \posA)$, associated with the used information, is solution to: 
\begin{gather}
            \label{201608221655}
            \fracp{\UsedGenerique}{\tempsA} (\tempsA, \posA) = \CoeffUtilisation \FixedGenerique  (\tempsA, \posA).
\end{gather}
\noindent The parameter $\DomaineIntSortieGenerique$ corresponds to the area of action of the circulating information on the \textit{Outcome}. $\SortieGenerique  (\tempsA)$ is the \textit{Outcome} of the model, given by:
\begin{gather}
            \label{201608221656}
                \SortieGenerique(\tempsA) = \int_{\DomaineIntSortieGenerique} \UsedGenerique(\tempsA, \posA) \, d\posA.
\end{gather}
\vspace{0.7cm}

\noindent In this mathematical model we imposed :
\begin{gather}
\forall \tempsA\in (0,+\infty),~\FluxForwardGenerique(\tempsA,0) = \FluxBackwardGenerique(\tempsA,0) \text { and }
\FluxBackwardGenerique(\tempsA,1) = \FluxForwardGenerique(\tempsA,1)
\end{gather}
\noindent These conditions allow the circulating information to move back and forth between the two edges of the domain.\\

The initial conditions $\FluxForwardGenerique(0,\posA)$, $\FluxBackwardGenerique(0,\posA)$, $\FixedGenerique(0,\posA)$, $\UsedGenerique(0,\posA)$ and $\SortieGenerique(0)$ are given for all $x$ in $(0,1)$.

\subsection{Applicability of the mathematical model and its different versions}
\label{Subsect:usage_Eq}

The previously presented mathematical model made of Equations (\ref{201608221651}), (\ref{201608221652}), (\ref{201608221653}), (\ref{201608221655}) and (\ref{201608221656}) can be used to simulate and predict an accumulative process. Hence, it can be used to study data relative to a total production over a given period.\par

The fourth equation of the model is the <<usage>> equation. This equation determines the action of the injected information on the variable to predict. Therefore, this equation has to adapt the different ways in which an intake or an injection may affect a biological variable.\\
To model a logistical growth, we added a limiter in this equation. In this case, the <<usage>> equation becomes:
\begin{gather}\tag{4b}
\fracp{\UsedGenerique}{\tempsA} (\tempsA, \posA) = \CoeffUtilisation \FixedGenerique  (\tempsA, \posA)\left(\frac{\Seuil - \SortieGenerique(\tempsA)}{\Seuil}\right)
 \label{eq:4b}
\end{gather}

\noindent With this version of the equation, data related to the change in weight of an animal can be tackled. This equation is essentially the differential equation of Verhulst \cite{10015246307}:
\vspace{-0.2cm}
\begin{gather}
\label{Verh}
\fracp{y}{t} (\tempsA) = r ~ y(t)\left(\frac{K - y(t)}{K}\right)
\end{gather}

\noindent whose structure is equivalent. Indeed, in the case when nothing depends on $x$, the value of $\CoeffUtilisation$ is very high and $\DomaineIntSortieGenerique$ is the whole interval $[0~;~1]$, $\UsedGenerique$, $\FixedGenerique$ and $\SortieGenerique$ are very similar. Hence, Equations (\ref{Verh})  and (\ref{eq:4b}) are essentially the same.\par
It may be also necessary to model variations to use our tool to treat, for example, data concerning drug effects. To do so, we have to be able to model an increase or a decrease in the \textit{Outcomes}. Since it is the case of most biological variables, we assumed that these outcomes may vary between an upper and a lower bound. Hence, we built two other equations: The equation,
\begin{gather}\tag{4c}
\displaystyle\fracp{\UsedGenerique}{\tempsA} (\tempsA, \posA) = -\Big(\UsedGenerique (\tempsA, \posA) - \Bsup \Big) -  \CoeffUtilisation \FixedGenerique  (\tempsA, \posA)\Big(\UsedGenerique (\tempsA, \posA) - \Binf \Big)
 \label{eq:4c}
 \end{gather}

\noindent models that the fixed information $\FixedGenerique$ orients the \textit{Outcomes} $\SortieGenerique$ toward a state that is lower than the steady state, and the equation,
\begin{gather}\tag{4d}
\displaystyle\fracp{\UsedGenerique}{\tempsA} (\tempsA, \posA) = -\CoeffUtilisation  \FixedGenerique  (\tempsA, \posA)\Big(\UsedGenerique (\tempsA, \posA) - \Bsup\Big) - \Big(\UsedGenerique (\tempsA, \posA) - \Binf \Big)
 \label{eq:4d}
 \end{gather}

\noindent models that the fixed information $\FixedGenerique$ orients the \textit{Outcomes} $\SortieGenerique$ toward a state which is greater than the steady-state. In these two cases, the \textit{Outcomes} vary between a lower bound $\Binf$ and an upper bound $\Bsup$.\\

The <<usage>> equation has to be defined a priori according to the variable to predict and the used inputs.\par
We expect that the mathematical models made of Equations (\ref{201608221651}), (\ref{201608221652}), (\ref{201608221653}), (\ref{201608221656}) and Equations (\ref{201608221655}),  \ref{eq:4b},  \ref{eq:4c} or  \ref{eq:4d} are sufficiently generic to be fitted on data relative to all the different farm species.\\

\subsection{Mesh and discretization of the Mathematical Model}
\label{Subsect:Mesh_Discret}

For the discretization of the {Mathematical Model}, we first used the classical Finite Difference method, with a given space step, to obtain semi-discrete in space equations. And, because the {Mathematical Model} is coded using R software, we used the R-function $Ode.1D$ developed by Soetaert et al. \cite{deSolvePackage} to manage the temporal discretization of the semi-discrete equations. This R-function calls upon the fourth-order Runge Kutta method with a given time step (See \cite{RungeKuttaOrdinaryDifferential}).\\
In this first exploration, to find a compromise between precision and calculation time, we parameterized the mesh with a time step of $0.001$ and a space step of $0.025$.

\subsection{The model parameters}

The system of PDEs contains several parameters : $\Velocity$, $\CoeffDiffGenerique$, $\CoeffTranstFToBGenerique$, $\CoeffFixGenerique$, $\CoeffUtilisation$ and $\Seuil$. The diffusion parameter $\CoeffDiffGenerique$ is the less influent model parameter. Hence, we set it to $0.001$. All the other parameters are fitted from a database by using an optimization algorithm. To do so, we used the function $directL$ developed by Johnson \cite{Steven2008}, which is embedded in R (\cite{RLanguage}) and applying the DIRECT algorithm developed by Finkel \cite{Finkel}. This algorithm searches the optimal values of the parameters, that is the values that minimize the error associated with the model on a given training database.\\

A detailed mathematical analysis of the model and its discretization will be performed in an upcoming paper. Nevertheless, we already know that the convection and diffusion speeds must follow the CFL conditions (See \cite{courant1928partiellen} and \cite{weisstein2014courant}). Indeed, since we set the discretization steps (Section \ref{Subsect:Mesh_Discret}), $\omega$ must be smaller than $25$ and $c$ must be smaller than $0.625$.\\

To fit the model we have to specify lower and upper values for each parameter between which the optimization algorithm will search their optimal values. Therefore, a comprehensive study of the ranges of values of the different model parameters was performed and presented in the working paper \cite{FlourentRangeValues}. We refer to it for the details of this study.

\section{Simulation tests of the learning capability of the model}
\label{sec:simulation}
The objective of this section is to present the tests by simulation performed to verify the ability of the tool to learn parameters from noisy biological data. To do so, we started by generating a fictitious database from the parameterized mathematical model made of Equations (\ref{201608221651}), (\ref{201608221652}), (\ref{201608221653}), (\ref{201608221655}) and (\ref{201608221656}). Then, we used this database to study the compensation effects existing between the parameters, to simulate the fitting of the parameters and then to verify if the model fits the data correctly.

\subsection{Generation of a fictitious database}

We generated a fictitious database containing 50 individuals, which is 50 $Output~Curves$. The objective was to obtain a database having the same characteristics as a real field database. To do so, we integrated noise and individual variations in it.\par
The construction of this fictitious database is presented in \hyperref[AppendI]{Appendix I}. We refer to it for the details.\par
Figure \ref{fig:ExampleCurves} shows an example of the generated curves without and with noise. We divided the obtained database into two parts: A {Training Database} made of $30$ curves and a {Test Database} made of $20$ curves.\par
In the rest of this section, we supposed that we have an experimental-like database and a model containing four parameter values to determine:  $\Velocity$, $\CoeffDiffGenerique$, $\CoeffTranstFToBGenerique$, $\CoeffFixGenerique$ and $\CoeffUtilisation$.
\begin{figure}[h]
    \centering
\includegraphics[width=0.55\textwidth]{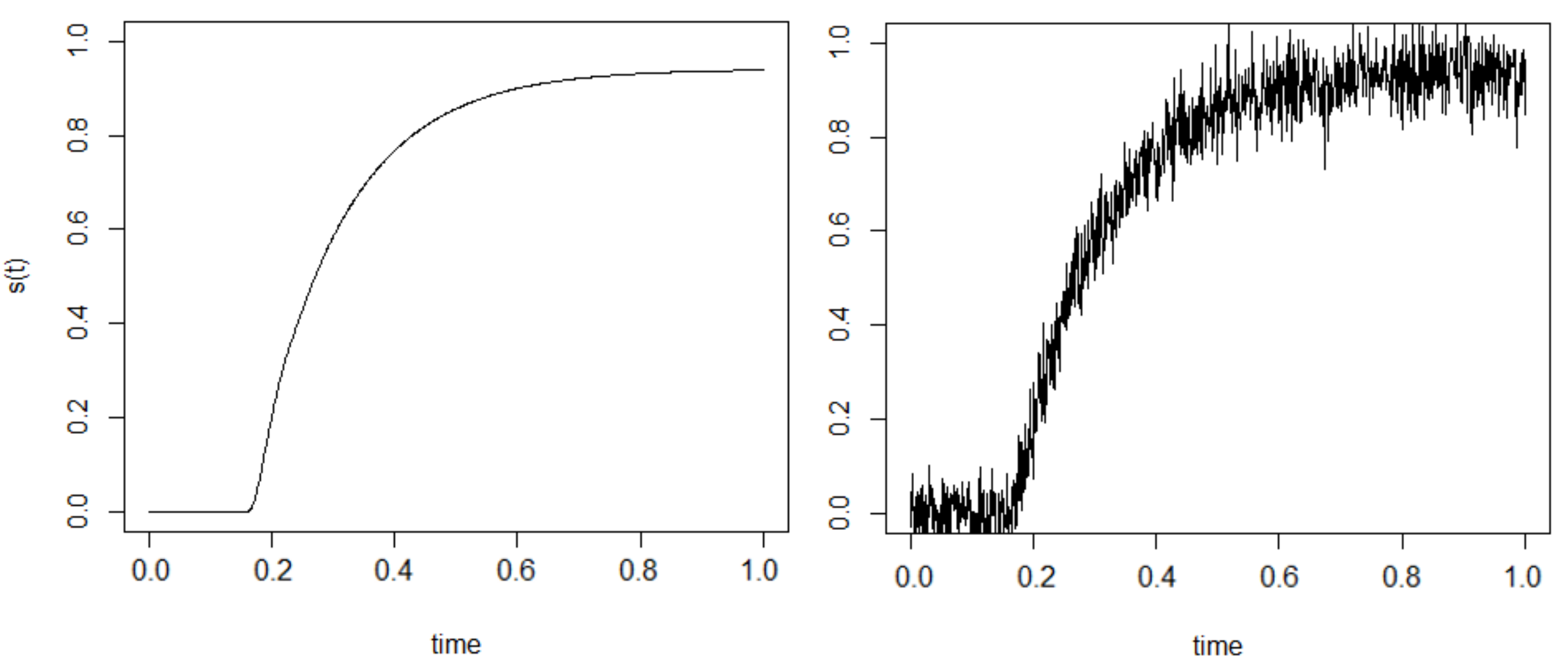}
    \caption{Example of simulated curves without and with noise}
    \label{fig:ExampleCurves}
\end{figure}

\subsection{Construction of relations linking some parameters}
\label{SubSection:Relation_param}

\begin{figure}[h]
    \centering
\includegraphics[width=0.8\textwidth]{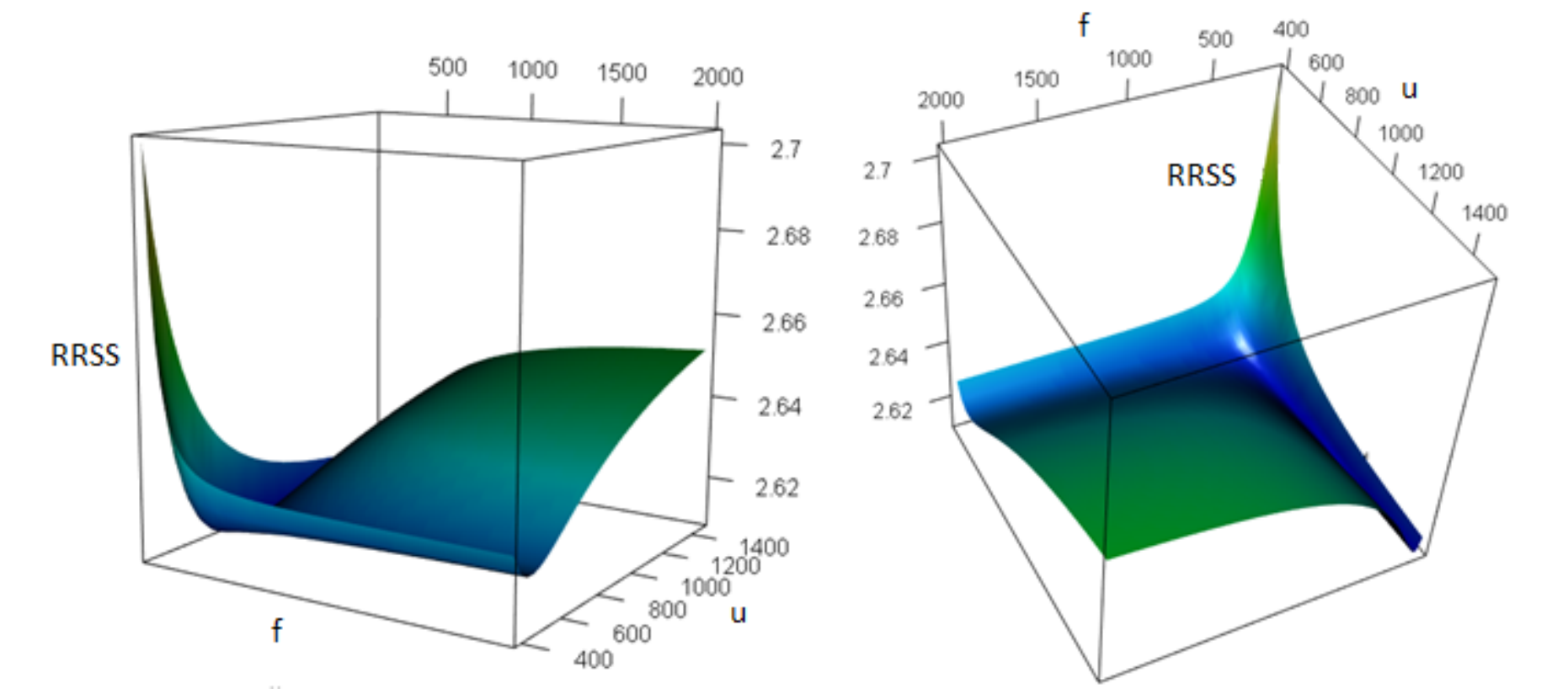}
    \caption{The $3D$ representation of the value of an error indicator (the Relative Residual Sum of Squares). This value is calculating according to the values of $\CoeffFixGenerique$ and $\CoeffUtilisation$.}
            \label{fig:RRSSfu3D_2}
\end{figure}

A study of the compensation effects existing between $\Velocity$ and $\CoeffTranstFToBGenerique$ and between $\CoeffFixGenerique$ and $\CoeffUtilisation$, put in evidence the existence of relations existing between these two pairs of parameters. For example, the relation existing between the parameters $\CoeffFixGenerique$ and $\CoeffUtilisation$ can be noticed in Figure (\ref{fig:RRSSfu3D_2}).

This study is presented in \hyperref[AppendII]{Appendix II}. We refer to it for the details.\\

We concluded from this study that, using a Nadaraya-Watson kernel regressions (See \cite{nadaraya1964estimating} and \cite{watson1964smooth}), we obtained a non-parametric relationship linking $\Velocity$ and $\CoeffTranstFToBGenerique$ in the form of,
\begin{gather}
\label{eq:model_r_w}
\CoeffTranstFToBGenerique= \hat{m_{\omega}}(\Velocity) + \epsilon_{\omega},
\end{gather}

\noindent and another one linking $\CoeffFixGenerique$ and $\CoeffUtilisation$ in the form of,
\begin{gather}
\label{eq:model_f_u}
\CoeffUtilisation= \hat{m_f}(\CoeffFixGenerique) + \epsilon_f,
\end{gather}

\noindent where $\hat{m_{\omega}}$ and $\hat{m_f}$ corresponds to the Nadaraya-Watson estimators and $\epsilon_{\omega}$ and $\epsilon_f$ are the residuals.\\

Knowing the relationship existing between $\Velocity$ and $\CoeffTranstFToBGenerique$ and the one existing between $\CoeffFixGenerique$ and $\CoeffUtilisation$ , it is possible to fit $\Velocity$ and $\CoeffFixGenerique$ and then deduce the values of $\CoeffTranstFToBGenerique$ and $\CoeffUtilisation$. Hence, these relations permits to reduce the number of parameters to learn simultaneously and so facilitate and reinforce the fitting process.

\subsection{Parameter fitting  and calculation of the model accuracy}
\label{Fitting_Parameters_Simul}

We fitted the parameters to the {Training Database} and then we tested the accuracy of the obtained model by calculating the error made on the {Test Database}.

\subsubsection{Fitting of $\Velocity$ and $\CoeffFixGenerique$ and accuracy of the obtained predictive tool}

To perform several fittings we sampled the {Training Database}: we sampled $20$ curves from the $30$ curves of the {Training Database} and we fitted the parameters to the $20$ sampled curves. By proceeding in this manner, we performed $30$ fittings of the parameters. To determine the values of $\Velocity$, $\CoeffTranstFToBGenerique$, $\CoeffFixGenerique$ and $\CoeffUtilisation$, we fitted $\Velocity$ and $\CoeffFixGenerique$ on the selected curves of the {Training Database} and we deduced the values of $\CoeffTranstFToBGenerique$ and $\CoeffUtilisation$ using Equalities (\ref{eq:model_r_w}) and (\ref{eq:model_f_u}).\par
To optimize the parameters, we used the R-function $directl$ to find the values of $\Velocity$ and $\CoeffFixGenerique$ minimizing the function, 
\begin{gather}
\label{fObj_wf}
\displaystyle{f_{obj} (\omega,f)= \frac{1}{n}\sum\limits_{i=1}^n\left(\sum\limits_{j=1}^m\left(\frac{(y_{ij_{obs}}-y_{ij_{pred}}(\omega,f))}{y_{ij_{obs}}}\right)^2\right)}.
\end{gather}

\begin{figure}[h!]
    \centering
\includegraphics[width=1\textwidth]{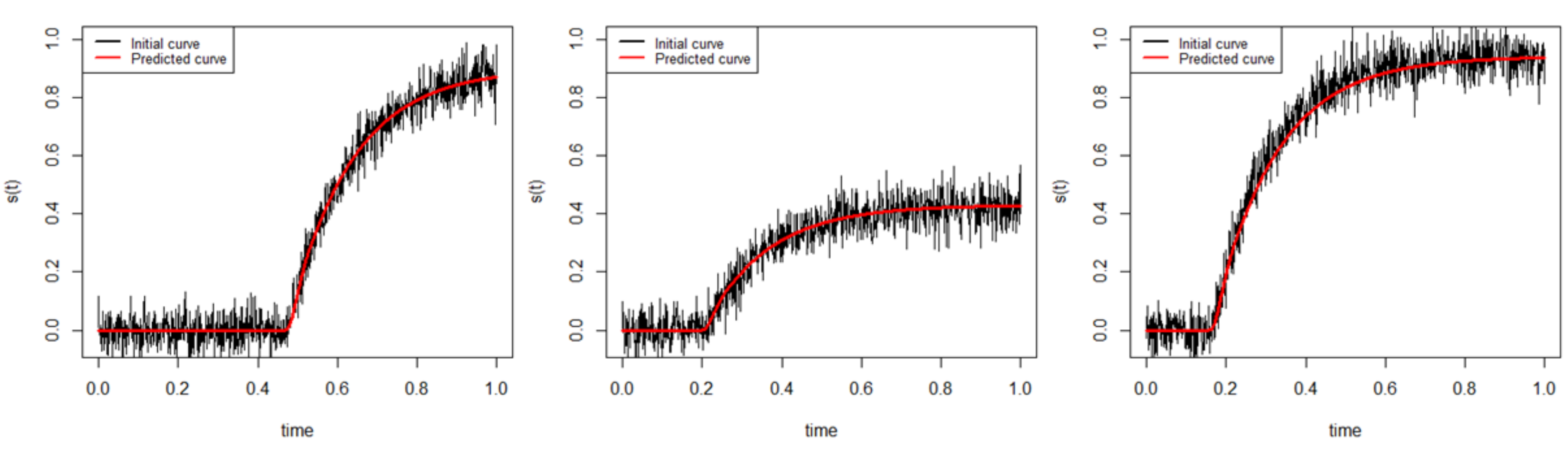}
    \caption{Examples of results given by the predictive tool in comparison with some training curves.}
        \label{fig:Examples_Simul}
\end{figure}

\noindent After the $30$ fittings of the parameters, we obtained $30$ values of $\Velocity$, $\CoeffTranstFToBGenerique$, $\CoeffFixGenerique$ and $\CoeffUtilisation$. We calculated the average value and the Relative Standard Deviation (RSD) of each parameter (Table \ref{table:distributions_simul}). We also looked at the fit of the model calculating from the {Training Database} the value of the Determination Coefficient ($R^2$) (Figure \ref{fig:Examples_Simul} and Table \ref{table:distributions_simul}). The results show that the model fits the curves of the {Training Database} well.\\

\vspace{-0.5cm}
\begin{table}[h!]
    \centering
            \caption{Average and Relative Standard Deviation of the parameters and the Determination Coefficient calculated on the {Training Database}.}
                \label{table:distributions_simul}
\begin{tabular}{|c|c|c|}
  \hline
  Parameter & Average & Relative standard deviation  \\
  \hline
  $\omega$ & $9.9$ & $0.009$  \\
  $f$ &  $920.3$ & $0.001$  \\
    $r$ & $35.6$ & $0.016$  \\
  $u$ &  $139.5$ & $0.001$  \\
  \hline
    $R^2$ &  $0.97$ & $0.011$  \\
  \hline
\end{tabular}
\end{table}

\begin{figure}[h]
    \centering
\includegraphics[width=0.9\textwidth]{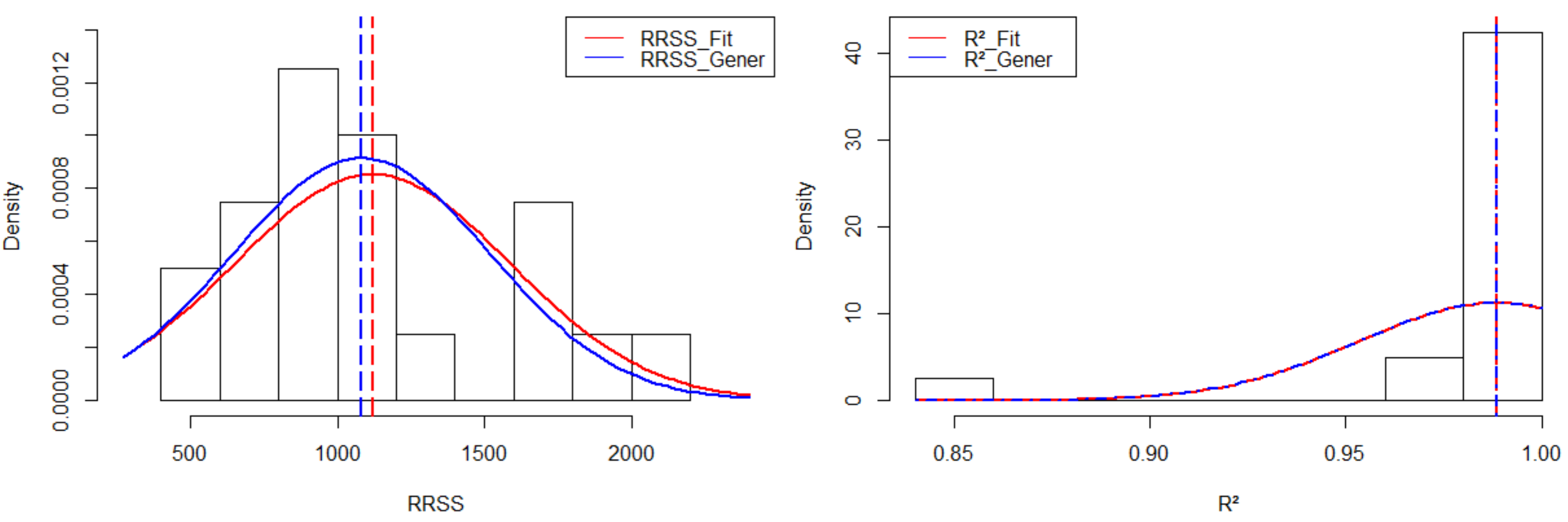}
    \caption{Distributions of the $RRSS$ and of the $R^2$ coefficient associated with the $Generator$ $Model$ and the {Fitted Model}.}
    \label{fig:Distrib_RRSS_R2_simul}
\end{figure}

To validate the ability of the tool to learn parameters from noisy data, we calculated the accuracy of the model on the {Test Database}. To do so, we calculated the Relative Residual Sum of Square ($RRSS$) and the Determination Coefficient associated with each curve contained in the {Test Database} and we obtained the distributions showed in Figure \ref{fig:Distrib_RRSS_R2_simul}. The $RRSS$ is low and the Determination Coefficient is high, indicating that the model fits the curves of the {Test Database} well.\\

We compared the $R^2$ and the $RRSS$ associated with the {Generator model} ($R^2_{Gener}$ and $RRSS_{Gener}$) - i.e. the model used to generate the fictitious Database - and the $R^2$ and the $RRSS$ associated with the {Fitted Model} ($R^2_{Fit}$ and $RRSS_{Fit}$) (Figure \ref{fig:Distrib_RRSS_R2_simul} and Table \ref{table:Comparison_simul}). $RRSS_{Fit}$ is low and this value is very similar to the value of $RRSS_{Gener}$. The $R^2_{Fit}$ is high and this value is also very similar to the value of $R^2_{Gener}$. These indicators thus demonstrate that the model fitting method is highly satisfactory and the error associated with the adjusted model is limited to the amount of noise and individual differences initially integrated into the generated database.

\vspace{-0.5cm}
\begin{table}[h]
    \centering
            \caption{Comparison between the indicators associated with the {Generator Model} and the {Fitted Model}.}
                \label{table:Comparison_simul}
\begin{tabular}{|c|c|c|}
  \hline
    ~ & $RRSS$ & $R^2$ \\
    \hline
  {Generator Model} &  $1082$ & $0.9887$\\
  \hline
    {Fitted Model}  & $1119$ & $0.9886$\\
  \hline
\end{tabular}
\end{table}

\section{Application of our Biomimetic Statistical Learning Tool on field data}
\label{Sec:Appli}

In this section, we present an application of our approach to field data. The database we used is confidential therefore only the dimensionless {Inputs} and {Outputs} are presented.

\subsection{Objectives of this application on field data}
The objective of this application is to build a tool that can predict the logistic growth of animals according to their initial weights and their intakes all along a given period.

\subsection{Adaptation of the basic model}
To mimic a logistic behavior, we used Equation (\ref{eq:4b}) as <<usage>> equation (Section \ref{Subsect:usage_Eq}). In this equation, $\Seuil$ corresponds to the maximum weight attainable by the animals of the studied species. Experts have an idea of the value of $\Seuil$. Therefore, during the model fitting, the value of this parameter was search in a restricted range of values.\par
Therefore, in this application we used the mathematical model made of Equations (\ref{201608221651}), (\ref{201608221652}), (\ref{201608221653}), (\ref{eq:4b}) and (\ref{201608221656}). This model considered as a growth model contained five parameters to fit: $\Velocity$, $\CoeffTranstFToBGenerique$, $\CoeffFixGenerique$, $\CoeffUtilisation$ and $\Seuil$.

\subsection{The data used}

The database we used is made of two parts corresponding to two different groups of animals monitored during two different periods (Table \ref{table:Descritpion_Data}). The first group contained $8$ individuals, monitored over a unit-period from $t = 0$ until $t = 1$. For this group, the weight of the animals was measured at $t = 0$ and at $t = 1$. The second group contained $7$ individuals, monitored from $t = 0$ until $t = 2.5$. For this group the weight of the animals was measured at $t = 0$, $t = 0.6$, $t = 1.52$ and at $t = 2.5$. For both groups, intakes of each individual were recorded over each time-step of $0.16$ time-unit. Therefore, for each individual, information relative to those intakes are periodically injected in the model with a time-step of $0.16$.\par
The dataset relative to the first group constitutes our {Training Database} and the dataset relative to the second group constitutes our {Test Database}.

\begin{table}[h]
    \centering
            \caption{Description of the data used.}
                \label{table:Descritpion_Data}
\begin{tabular}{|c|c|c|}
  \hline
    ~ & First group & Second group\\ \hline
  Number of individuals & $8$   &  $7$\\  \hline
 ~ & $t=0$  & $t=0$\\ 
 Output measured at  & $t=1$ & $t=0.60$ \\ 
  ~  & ~ & $t=1.52$\\ 
  ~   & ~ & $t=2.50$ \\  \hline
  Time step of the \textit{Entries} injections & $\Delta t_{In}=0.16$   & $\Delta t_{In}=0.16$ \\  \hline
\end{tabular}
\end{table}

\subsection{Parameter fitting}

As in Section \ref{SubSection:Relation_param} and \hyperref[AppendII]{Appendix II}, we built a relationships between some parameters of the model by applying the same methodology. Using a Nadaraya-Watson kernel regressions, we obtained a non-parametric relation linking $\Velocity$ and $\CoeffTranstFToBGenerique$ and another one linking $\CoeffFixGenerique$ and $\CoeffUtilisation$. Knowing the relationships between these parameters, it is possible to fit $\Velocity$ and $\CoeffFixGenerique$ and then deduce the value of $\CoeffTranstFToBGenerique$ and $\CoeffUtilisation$. Therefore in this application we only fitted  $\Velocity$, $\CoeffFixGenerique$ and $\Seuil$ and deduced the values of $\CoeffTranstFToBGenerique$ and $\CoeffUtilisation$.\par 
The parameters were fitted to the {Training Database} by minimizing the difference between the simulated and the real {Outputs} at time $t = 1$. Hence, to fit the parameters, we used the algorithm DIRECT that minimized the function,
\begin{gather}
\label{eq:fobj_Application}
f_{obj}(\Velocity,\CoeffFixGenerique,\Seuil)= \frac{1}{n}\sum\limits_{i=1}^n\left(\frac{(s_{i_{obs}}(1)-s_{j_{pred}}(1))}{s_{i_{obs}}(1)}\right)^2,
\end{gather}

\noindent where $n$ is the number of individuals contained in the training database and $O_{i_{obs}}(1)$ and $O_{i_{pred}}(1)$ correspond respectively to the values of the observed and the predicted {Outputs} values for the $i^{th}$ individual at $t = 1$. \\

To performed several fitting procedures. We sampled the {Training Database}: we randomly selected $7$ individuals from the $8$ individuals before each fitting procedure and we fitted the parameters on the data associated with the selected individuals. Therefore, we performed $8$ fittings and we obtained $8$ sets of values of $(\Velocity, \CoeffTranstFToBGenerique, \CoeffFixGenerique, \CoeffUtilisation, \Seuil)$.

\subsection{Results}

We calculated the average and the $RSD$ of each parameter (Table \ref{table:Res_Application}). The $RSD$ of each parameter is low, indicating that our fitting method permitted to identify one set containing the parameter values that minimize the error associated with the {Fitted Model}. The existence of a single optimal set of values of $(\Velocity, \CoeffTranstFToBGenerique, \CoeffFixGenerique, \CoeffUtilisation, \Seuil)$ attests to the identifiability of the model.\\

We parameterized the model with the average values of the parameters.\par
We calculated the error associated with the model on the {Training Database}. To do so, we calculated the Average Relative Error ($ARE$) between the measured and predicted values of the {Output} at time $t = 1$:

\begin{gather}
\label{eq:err_Application}
ARE(t)= \frac{1}{n}\sum\limits_{i=1}^n\sqrt{\left(\frac{(s_{i_{obs}}(t)-s_{j_{pred}}(t))}{s_{i_{obs}}(t)}\right)^2}
\end{gather}

\begin{table}[h!]
    \centering
            \caption{Average values and $RSD$ of the fitted parameters. $ARE$ calculated at time $t = 1$ on the {Training Database}.}
                \label{table:Res_Application}
\begin{tabular}{|c|c|c|}
\hline 
$Parameter$ & $Mean$ & $RSD$ \\ 
\hline 
$\omega$ & $9.24$ & $0.079$\\ 
$r$ & $17.91$ & $0.14$ \\ 
$f$ & $707.01$ & $0.36$ \\ 
$u$ & $21.49$ & $0.17$ \\ 
$\Seuil$ & $1.70$ & $0.009$ \\ 
\hline 
$ARE(1)$ $(\%)$ & $1.83$ & $0.013$ \\ 
\hline 
\end{tabular}
\end{table}

The $ARE$ value calculated on the {Training Database} at time $t = 1$, is worth $1.83\%$. This result is satisfactory, but the accuracy of the model must be calculated on a {Test Database} to ensure that the model does not overfit the training data. \\

\begin{figure}[h!]
	\centering
\includegraphics[width=1\textwidth]{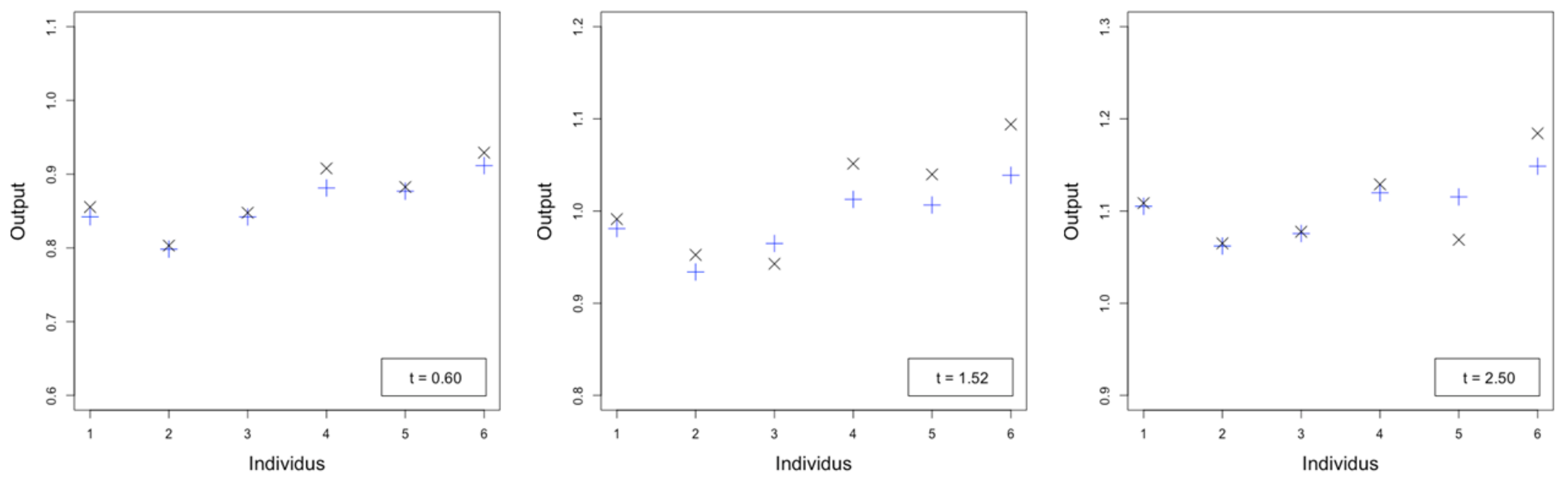}
	\caption{Difference obtained between the measured ($+$) and predicted ($\times$) values of the {Output} variable at different times $t$ for the individuals in the {Test Database}.}
	\label{fig:res_BaseTest}
\end{figure}

To do so, we calculated the $ARE$ given by Equation (\ref{eq:err_Application}), on the {Test Database} at time $t = 0.6$, $t = 1.52$ and $t = 2.5$ (Table \ref{table:Res_ApplicationTest} and Figure \ref{fig:res_BaseTest}).

\begin{table}[h!]
	\centering
    		\caption{Average Relative Error ($ARE$) calculated on the {Test Database} at different instants.}
				\label{table:Res_ApplicationTest}
\begin{tabular}{|c|c|c|c|c|}
\hline 
t & $0.6$ & $1.52$ & $2.5$ \\ 
\hline 
ARE(t) (\%) & $1.3$ & $2.9$ & $1.5$ \\ 
\hline 
\end{tabular}
\end{table}

\subsection{Discussion of the results}

The error associated with the model is low on the {Test Database}. The errors made before and after time $t = 1$ remain low. Those results indicate that our tool can be trained on a very small database to link the inputs and the outputs and then it can accurately predict the weight of the animals over a period $2.5$ times longer than the training one.\\
This extrapolation capability, obtained despite the very low quantity of training data, illustrates that our tools hold high potential for information extraction. As it will be demonstrated below, this capability distinguishes our approach from other inference methods.\par
Moreover, in addition to the information extraction potential, the extrapolation capability helps to reduce the training-data-dependency of our tool. Indeed, we demonstrated that our tool can be applied outside the training data range and provide accurate extrapolations. Hence, we do not need to fit it on data covering the whole curve to predict and so we can use smaller {Training Database}. Therefore, this extrapolation capability permits to reduce the duration of data collection, the duration of in situ experiments, and thus the computational and the experimental costs.

\section{Comparison with existing growth models}

According to \cite{VazquezCruz2014} and \cite{EvolutionaryAlgorithms}, the current methods used to simulate and predict logistic growth processes, involve two main types of models: Phenomenological Models corresponding to <<Black Box>> models, and Mechanistic Models corresponding to <<White Box>> models. In this section, we will compare some models belonging to these two main categories with the Biomimetic Statistical Learning tool presented in this paper.

\subsection{The Phenomenological Models}

 As defined in \cite{VazquezCruz2014}, the Phenomenological Models include Linear, Multiple Linear and Nonlinear Regressions, Logistic Models and Neuronal Networks.

\subsubsection{Comparison of our Biomimetic Growth Model with Classical Logistic Growth Models}
The models of Gompertz \cite{gompertz1825xxiv}, 
\begin{gather}
\label{eq:Gompertz}
\frac{dN(t)}{dt}= a_G.N(t). \ln\left(\frac{K_G}{N(t)}\right),
\end{gather}
\noindent and Verhulst \cite{10015246307},
\begin{gather}
\label{eq:Verhulst}
\frac{dN(t)}{dt}= a_V.N(t). \left(1 - \frac{N(t)}{K_V}\right),
\end{gather}
\noindent are two models frequently used to model growth processes (e.g. see \cite{Winsor1}, \cite{sakomura2005modeling}, \cite{buyse2004assessment}, \cite{robertson1916experimental}, \cite{robertson1923chemical} and \cite{roman2012modelling}).\\
We fitted the parameters of these two models on our {Training Database} by using the same optimization algorithm that we used to fit the Biomimetic Model and by minimizing $ARE(1)$.\\ 

\begin{figure}[htp]
\begin{minipage}{0.5\textwidth}
	\includegraphics[width=0.9\textwidth]{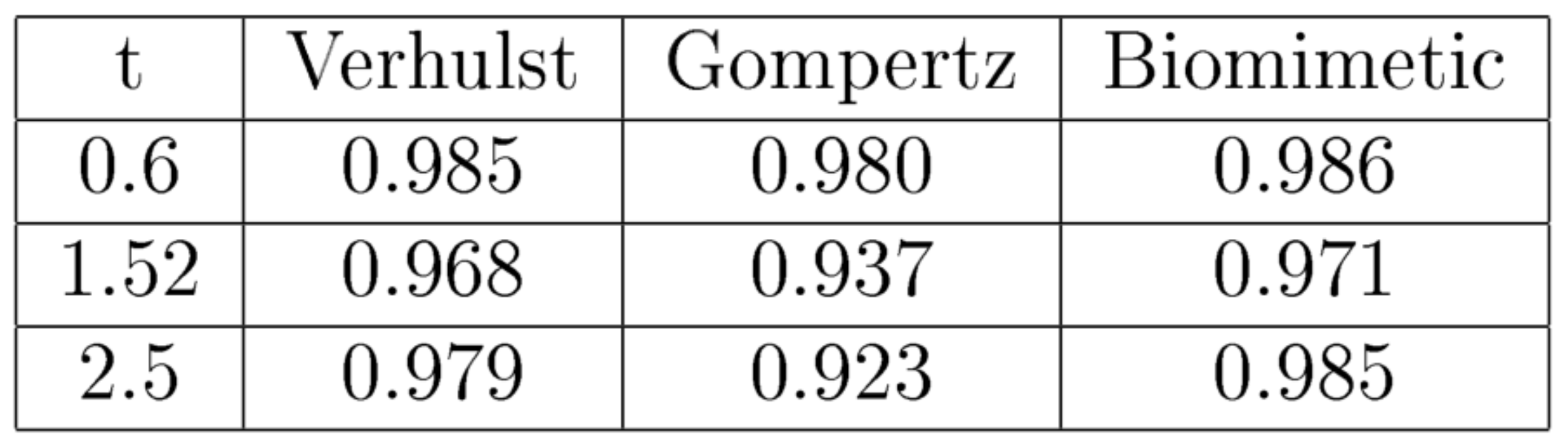}
\end{minipage}
\begin{minipage}{0.5\textwidth}
		\centering
\includegraphics[width=0.8\textwidth]{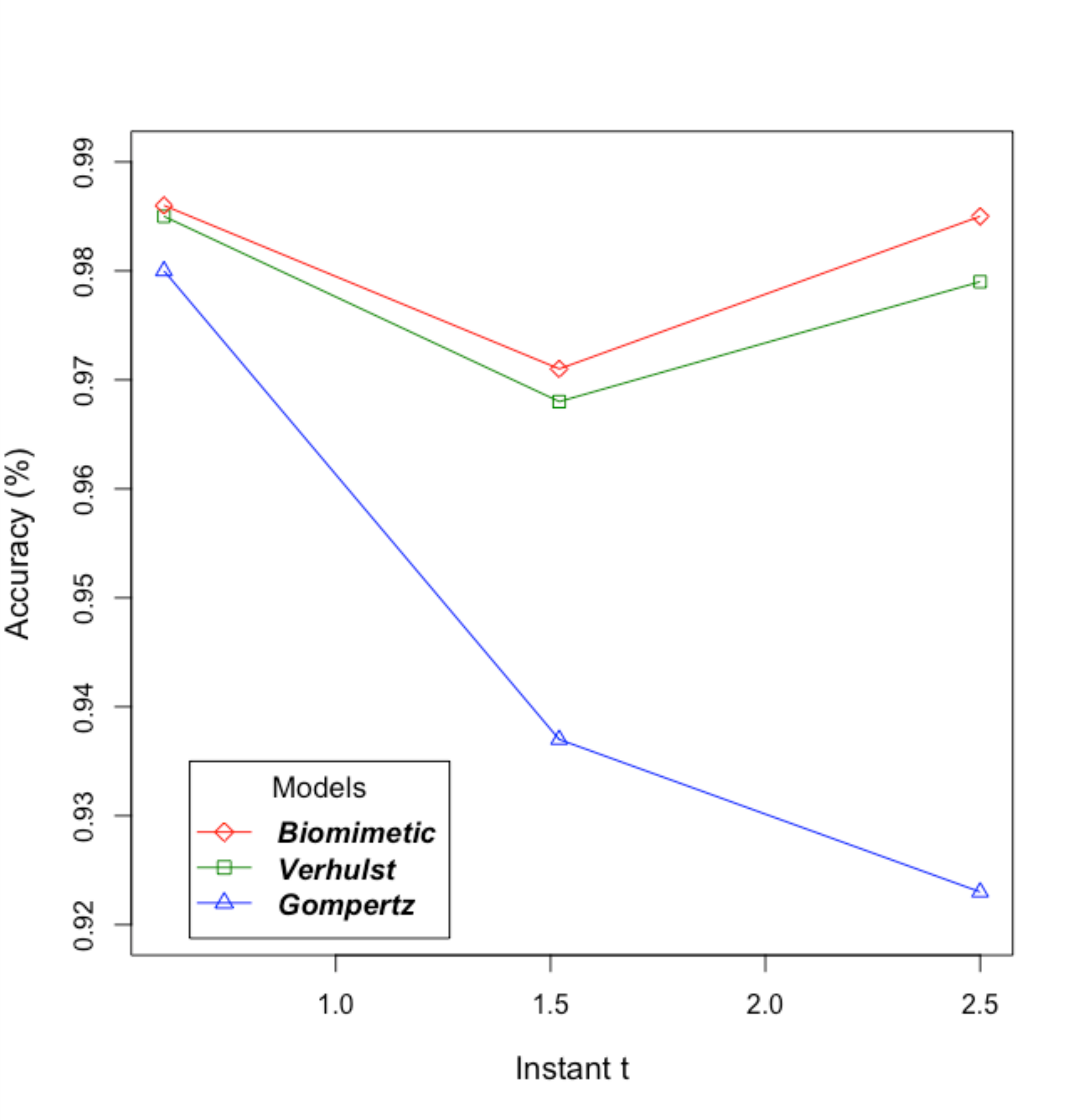}
\end{minipage}
  \caption{The $ARA$ calculated on the {Test Database} at different times $t$ and associated with different models.}
	\label{fig:Accu_Class_Mod_figure}
\end{figure}

\begin{figure}[h!]
	\centering
	\includegraphics[width=0.75\textwidth]{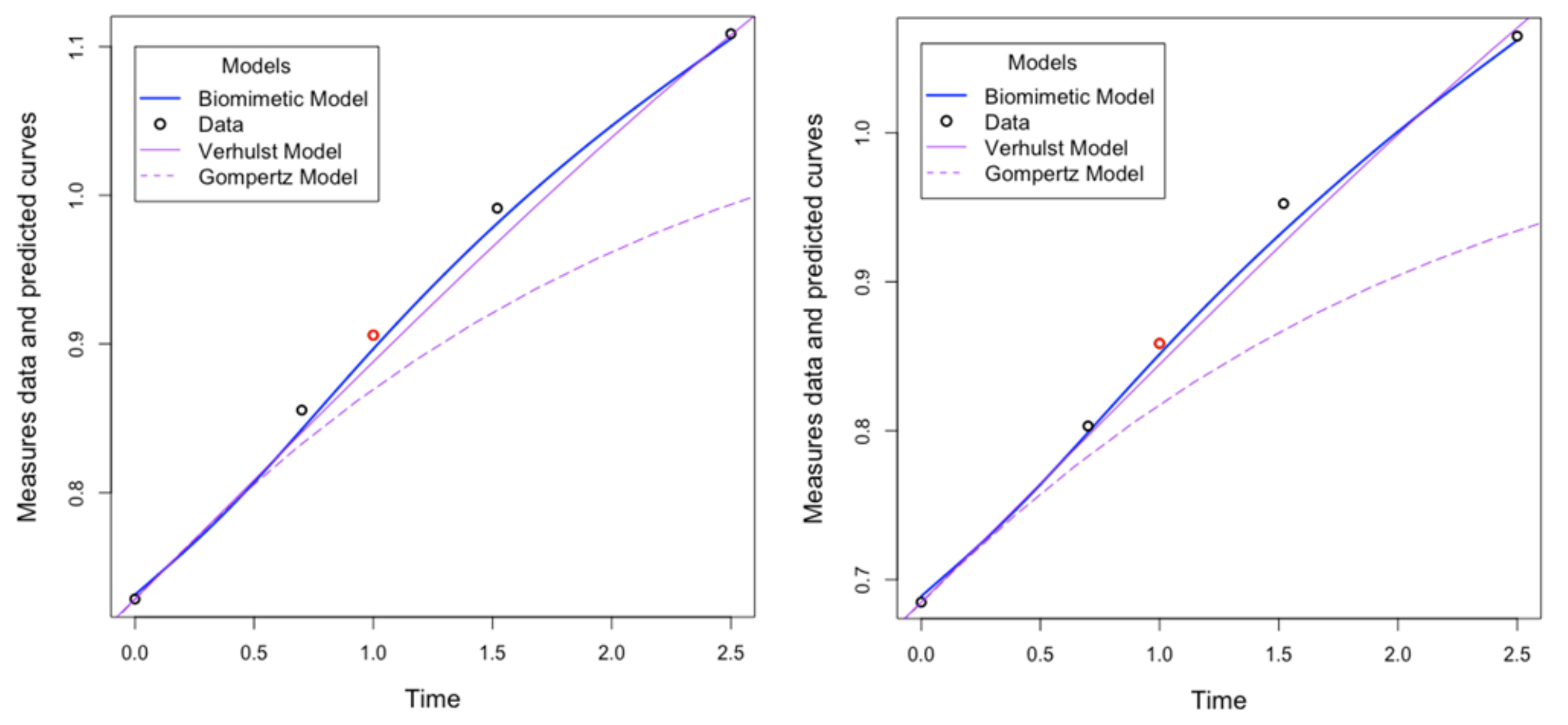}
	\caption{Plot of the predicted growth curves of two individuals contained in the {Test Database} with the different models.}
	\label{fig:Accu_Class_Mod}
\end{figure}

To test and compare the accuracy of the different models, we calculated on the {Test Database} the Average Relative Accuracy, $ARA$ at different times $t$ : 
\begin{gather}
\label{eq:ARA}
ARA(t) = 1-ARE(t),
\end{gather}

\noindent where $ARE$ is given by Equation (\ref{eq:err_Application}).
\noindent We used the three parameterized models to generate the growth curve of the individuals of the {Test Database} and we compared the measured and the predicted values at times $t = 0.6$, $t = 1.52$ and $t = 2.5$.\\
The results contained in Table \ref{table:Param_classic_mod} and the curves of Figures \ref{fig:Accu_Class_Mod_figure} and \ref{fig:Accu_Class_Mod} show that the curves generated from the Gompertz's model featured a premature deceleration. However, the Verhulst's model is associated with good accuracy over the whole studied period.\par
The similarity of the results from the Verhulst and the Biomimetic Growth Models was expected because our model includes an equation assimilable to the Verhulst's equation (see Section \ref{Subsect:usage_Eq}). The real advantage of our biomimetic growth model is its ability to integrate input data. Indeed, the Verhulst equation only takes into account the initial conditions of the system under study, whereas our model also integrates intakes throughout the studied period. The capability to integrate additional information appears to help refine the results and increase the accuracy of our model. Moreover, since the Verhulst's and the Gompertz's models can not integrate input data over time, they are not able to perform Data Assimilation, contrary to our tool.

\begin{table}[h]
	\centering
    		\caption{Parameter values and $ARA(1)$ calculated on the {Training Database}.}
				\label{table:Param_classic_mod}
\begin{tabular}{|c c c c|}
\hline 
	Model & a & K & ARA(1) \\  \hline
	Gompertz & $a_G=0.412$ & $K_G=0.563$ & $0.978$ \\ 
	Verhulst & $a_V=0.411$ & $K_V=1.563$ & $0.979$  \\  \hline
	Biomimetic &   &     & $0.981$ \\ \hline
\end{tabular}
\end{table} 

\vspace{-0.5cm}

\vspace{0.2cm}
\subsubsection{Comparison between the Biomimetic Growth Model and Neural Networks}

Over the past decade, the use of Machine Learning (ML) algorithms and especially Neural Networks (NN) has been on the rise \cite{DomingosLearning}. According to some studies (\cite{GORCZYCA2018286}, \cite{VALLETTA2017203}, \cite{MA2014798} and \cite{IP2018376}), the popularity of these tools can be explained by the ease of their implementation and the diversity of issues that these algorithms can handle. Nevertheless, these algorithms are based on relatively simple mathematical models that are cannot easily take into account complex phenomena, such as delay and saturation. Hence, we applied different Neural Networks on our {Training Database} to compare this kind of ML tool and our Biomimetic Growth Model. We tested six Neural Networks having different numbers of nodes and hidden layers, taken as inputs the initial weight of each individual and their periodically recorded intakes (Table \ref{table:ARE_NN}).

\begin{table}[h!]
	\centering
    		\caption{The $ARA$ calculated on the {Training Database} ($ARA_{Train}$), and on the {Test Database} ($ARA_{Test}$), at $t=1$, with different Neural Networks. The Neural Network ($k_1$,...,$k_i$,...,$k_n$) corresponds to a Neural Network containing $n$ hidden layers and the $i^{th}$ hidden layer contains $k_i$ nodes.}
				\label{table:ARE_NN}
\begin{tabular}{ | c | c | c |}
\hline
	Structure & $ARA_{Train}(1)$ ($\%$)  & $ARA_{Test}(1)$ ($\%$)  \\ \hline
	(4) & 99.9 & 78.8 \\ \hline
	(4,3) & 99.8 & 90.5 \\ \hline
	(6,5) & 99.7 & 93.4 \\ \hline
	(4,6,6,3) & 99.9 & 94.8 \\ \hline
	(5,7,7,7,4) & 99.8 & 95.3 \\ \hline
	(5,9,9,9,5) & 99.9 & 93 \\ \hline
\end{tabular}
\end{table}

We fitted each tested Neural Network on our {Training Database} by using the R-function $neuralnet$ developed by Fritsch et al. \cite{NeuralnetPackage}, and we calculated the accuracy of those Neural Networks on the $Training$ and on the {Test Database}.\par
The results given in Table \ref{table:ARE_NN} show that all the tested Neural Network fit the curves of the {Training Database} better than the ones of the {Test Database}. It shows that the tested Neural Network overfit the training curves, particularly when the structure of the studied Neural Networks is composed of too many or too few nodes and hidden layers. The accuracy of the Neural Networks on the {Test Database} increases up to a certain number of nodes and hidden layers and then decreases when the complexity of the structure continues to increase. On the test database, the highest accuracy value is reached using a Neural Networks containing $5$ hidden layers, but this value is lower than that obtained using our Biomimetic Model (Table \ref{table:ARE_NN} and Figure \ref{fig:Accu_Class_Mod_figure}).\par
Nevertheless, the accuracy of these ML tools is satisfactory and the real advantage of our Biomimetic tool over Neural Networks is its extrapolation capability. Indeed, as the Biomimetic Model, the studied Neural Networks were fitted only from the value of the {Output} at $t=1$. In this case, the fitted classical Neural Networks can only be used to predict the {Output} at $t=1$. Hence, Neural Networks cannot interpolate or extrapolate, in contrast to our Biomimetic Model. Therefore, contrary to our tool, Neural Networks can not be used in a "Biological Small Data" context to reduce the experimental and computational costs. They are also less suitable to perform Data Assimilation in this context.

\subsection{Mechanistic Growth Models}

Mechanistic Growth Models are another kind of tool permitting to gathered biological inputs to predict the growth of plants or animals. Some models of this type have been developed in \cite{Bastianelli1}, \cite{articleMach}, \cite{articleLafleur} and \cite{CESARTREJOZUNIGA2014474}. These models integrate numerous {Inputs}, and not all of which are available in our study. Hence, these models can not be applied to our database. Therefore, we only compared the structure, the functioning and the objectives of those Mechanistic Models with our Biomimetic one.\par
As it is said in \cite{VazquezCruz2014}, \cite{TEDESCHI2005}, \cite{Bastianelli2} and \cite{BEEVER1991115}, the construction of Mechanistic Growth Models generally focuses on the biological meaning of the overall model. Therefore, the construction of the explanatory mechanistic models takes time, requires a large quantity of zootechnical knowledge and results in complex models. As it is explained in \cite{wallach2001parameter}, \cite{Bastianelli2} and \cite{emmans1995problems}, these models contain a large number of unknown parameters and include many factors, forcing the user to enter a large number of {Input} values, which are sometimes difficult or costly to obtain. Hence, the complex structure of these models makes Mechanistic Realistic Models inappropriate for fitting data and Data Assimilation.\par

\section{Conclusion}

To conclude, we built a \textit{Biomimetic Statistical Learning tool} based on a PDE system embedding the mathematical expression of biological determinants. This PDE system contains parameters that can be fitted to data. This PDE system was carefully designed to have a high learning potency and a great accuracy but also to remain light and flexible.\par 
In the particular <<Biological Small Data>> context, the performed applications showed that this tool has higher accuracy than the existing tools. However, our \textit{Biomimetic Statistical Learning tool} distinguishes itself in light of its suitability to perform Data Assimilation even if the frequency of data collection and the quality of the collected data are low.\\
The extrapolation capability of our tool, coupled with its high learning potency permits to fit it on a few data but also to accurately applied it outside the range of the training data. Hence, the quantity of collected data can be reduced as the costs relative to experiments and data management.\par
To sum up, our tool can be used to predict health and performance indicators according to the ingestion or the injection of molecules in an animal body and to perform accurate and inexpensive Livestock Data Assimilation .\\
The pursuit of an optimal combination between the use of data and the use of prior knowledge via the use of PDEs seems to be an interesting way to build Artificial Intelligence (AI) tools. Those AI tools could have a strong learning ability and a weak Training-Data-Dependency.\par
Nevertheless, the results coming from the Biomimetic Model was obtained from a certain number of hypotheses. Some Model Selection methods could be applied to select the structure of the {Mathematical Model}, permitting to obtain a more satisfying model in terms of $ARE$ and the number of parameters to learn. This suggested improvement will be studied in a forthcoming work.\\

\section*{Acknowledgements}

The authors are very grateful to D. Causeur, G. Durrieu and E. Fokou\'{e} for fruitful discussions related to this article.

\printbibliography


\newpage
\setcounter{section}{1}
\renewcommand\thesection
{\Roman{section}}

\section*{Appendix I: Generation of the Learning Database}
\label{AppendI}

To test the learning capability of the model we generated a Learning Database containing 50 individuals, that is 50 $Output~Curves$. The objective is to obtain a database having the same characteristics as a real field database. To do that we integrated into this fictitious database noise and individual variability.

\subsection{Integration of individual variability}

\begin{figure}[h]
    \centering
    \includegraphics[width=0.65\textwidth]{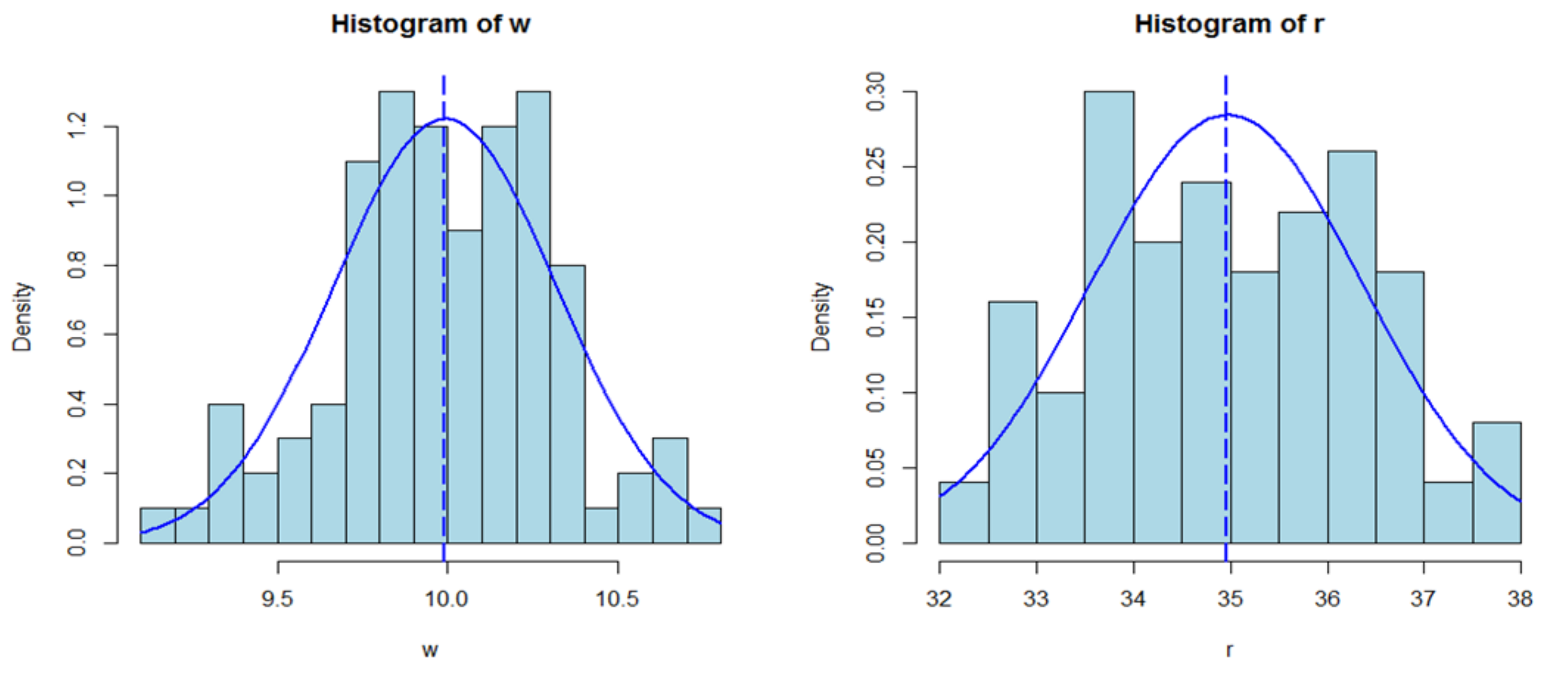}\\
\includegraphics[width=0.65\textwidth]{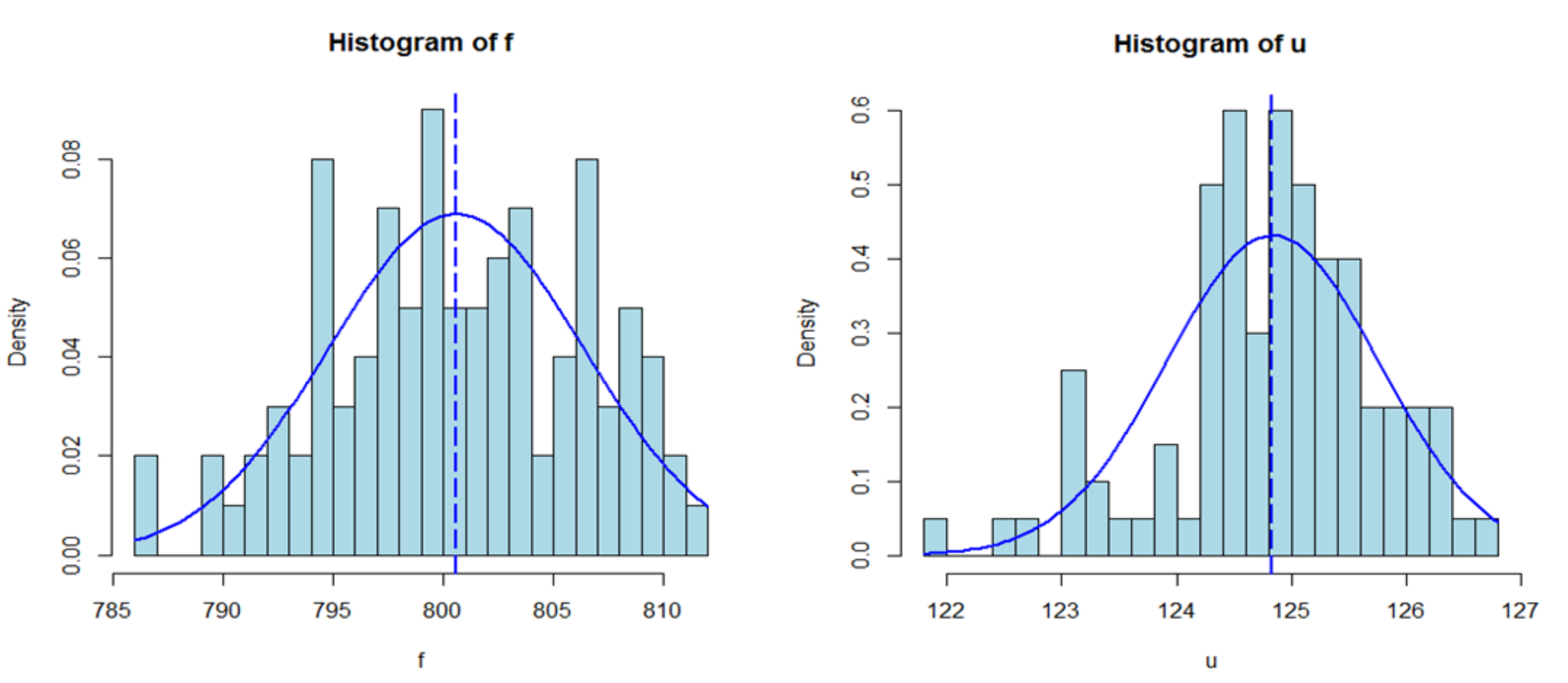}
    \caption{Distributions of the parameters $\omega$, $r$, $f$ and $u$}
    \label{fig:distributions}
\end{figure}

The model parameters are constants to determine. Nevertheless, to introduce individual variability in the generated data, we considered (only in this Section) the parameters as biological-like factors following a Normal distribution. To simulate individual differences, we assigned to each parameter a Normal distribution centered on an arbitrarily chosen value and with a relative variance of $0.005$ (See Table \ref{table:distributions}). From these Normal probability laws, we generated $50$ values of the parameters $\Velocity$, $\CoeffTranstFToBGenerique$, $\CoeffFixGenerique$ and $\CoeffUtilisation$. Their respective statistical and probabilistic distributions are given in Figure \ref{fig:distributions}.

\subsection{Generation of fictitious {Inputs}}

The $Inputs$ integrated into the model correspond to the injected volume ($VolQ$) and the moment of the injection ($c_t$). These parameters can take on any value between $0$ and $1$, therefore we applied a Uniform distribution over the interval $[0 ; 1]$ to these two types of $Inputs$ (Table \ref{table:distributions}).\\

From the values of the parameters and the fictitious $Inputs$, we generated $50$ $Output~Curves$. 

\subsection{Addition of a random noise}

Continuing with the objective of obtaining an experimental-like database, we added noise to the $Output~Curves$. To do so, we added a random component following a Gaussian distribution centered on $0$ and with a variance of $0.05$  to the generated curves (Table \ref{table:distributions}).\\

Figure \ref{fig:ExampleCurves} shows some examples of generated curves without and with noise. We divided the obtained database into two datasets: A $Training~Database$ made of $30$ curves and a $Test ~Database$ made of $20$ curves.\par
In the rest of this Section, we assumed that we have an experimental-like database and a model containing four parameter values to determine.

\begin{table}[h]
    \centering
            \caption{The distributions followed by the parameters and the $Inputs$.}
                \label{table:distributions}
\begin{tabular}{|c|c|}
  \hline
  Parameter & Probability law  \\
  \hline
        $\omega$ & $\mathcal{N}(10,0.3125)$  \\
        $r$ & $\mathcal{N}(35,1.42)$  \\
        $f$ & $\mathcal{N}(800,5.175)$  \\
        $u$ & $\mathcal{N}(125,1)$  \\
        $VolQ$ & $\mathcal{U}(0,1)$  \\
        $c_t$ & $\mathcal{U}(0,1)$  \\
        $Noise$ & $\mathcal{N}(0,0.05)$  \\
  \hline
\end{tabular}
\end{table}

\begin{figure}[h]
    \centering
\includegraphics[width=0.6\textwidth]{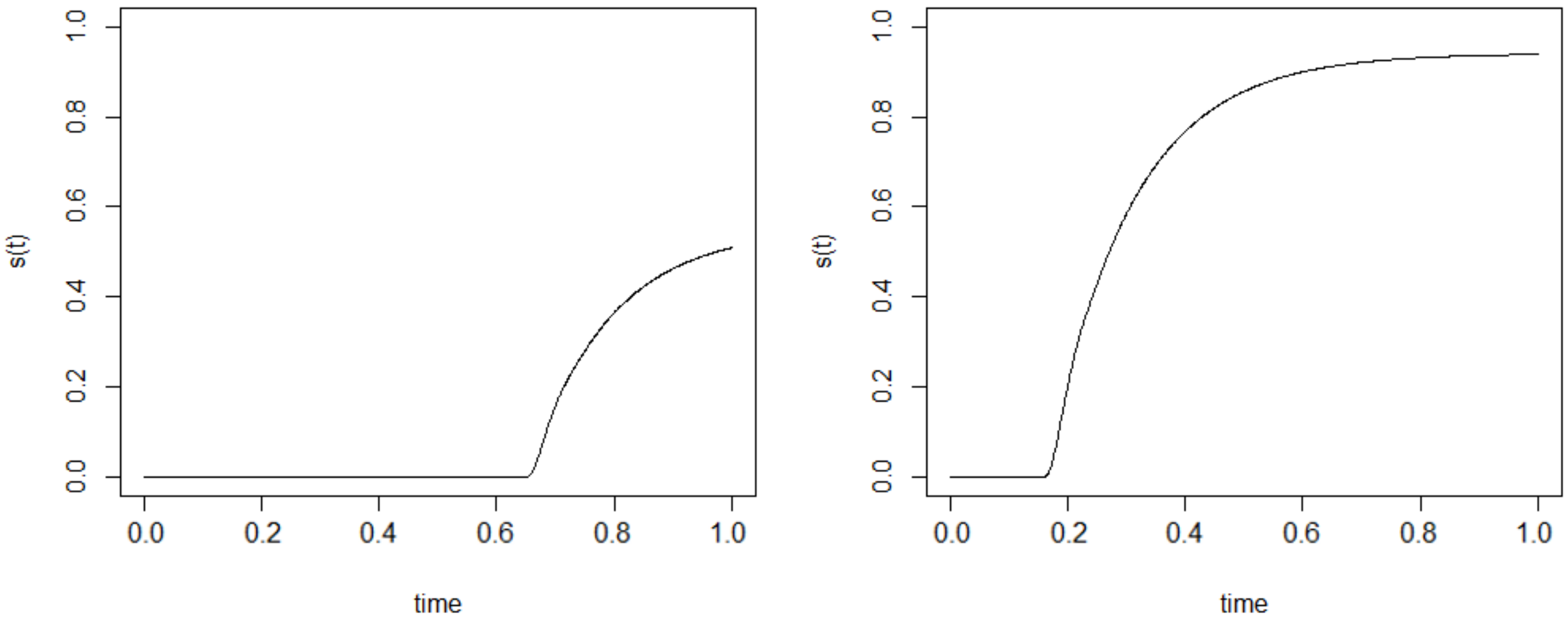}
\includegraphics[width=0.6\textwidth]{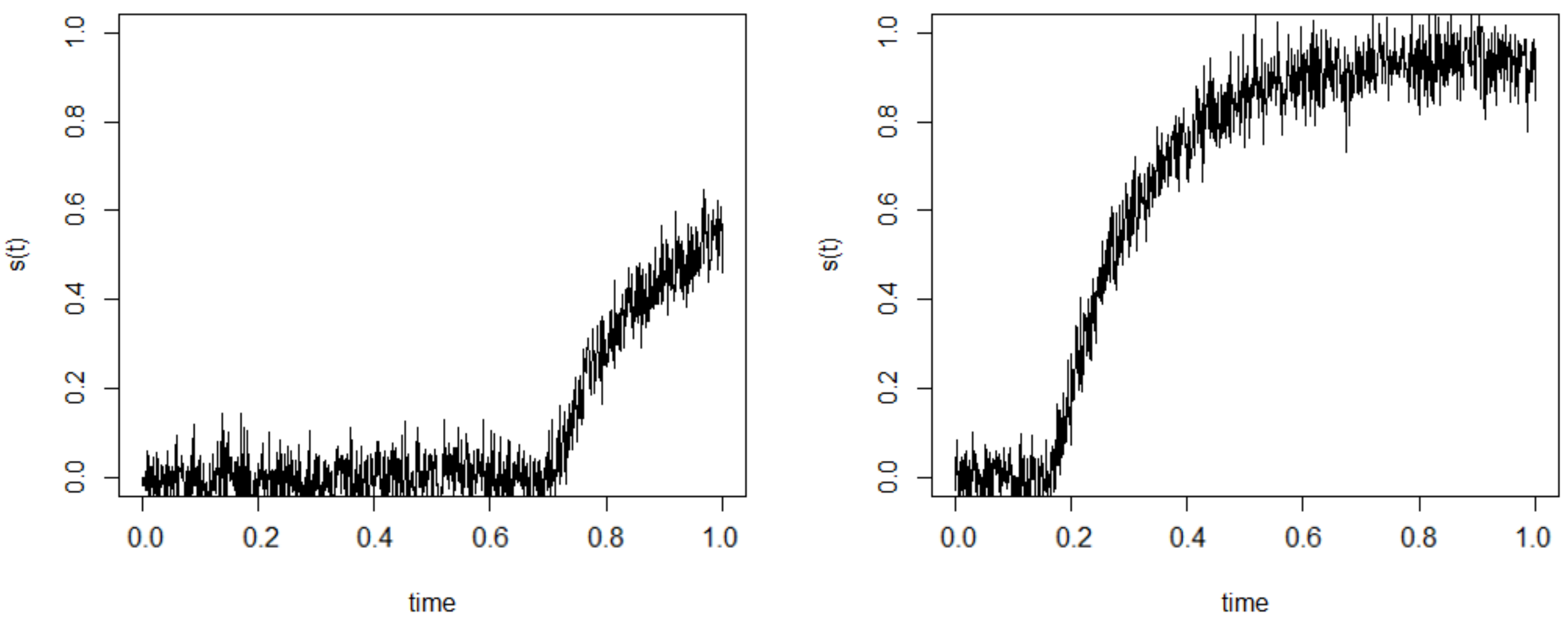}
    \caption{Example of simulated curves without and with noise}
    \label{fig:ExampleCurves}
\end{figure}


\newpage
\setcounter{section}{2}
\setcounter{subsection}{0}
\renewcommand\thesection
{\Roman{section}}

\section*{Appendix II: Study of the compensation effects existing between the model parameters}
\label{AppendII}

Among the parameters $\Velocity$, $\CoeffTranstFToBGenerique$, $\CoeffFixGenerique$ and $\CoeffUtilisation$, some parameters offset each other.\par
Velocity $\Velocity$, can be offset by any delay $\CoeffTranstFToBGenerique$, the information undergoes. For example, a low convection speed associated with a low delay may induce kinetics equivalent to that induced by a high convection speed associated with a long delay.\par
The fixation $\CoeffFixGenerique$, and the use of the information $\CoeffUtilisation$, are also two counterbalanced processes. For instance, a high fixation rate followed by a low usage of the information can induce the same effect on the $Outcome$ as a low fixation rate followed by an important use of the fixed information.\\

Therefore, relations exist between the parameters of those two couples. The objective of this part is to use the fictitious $Training~Database$ to study these relations.\\

\subsection{Study of the relationship between $\Velocity$ and $\CoeffTranstFToBGenerique$}
\label{Subsection:Relation_w_r}

First, we demonstrated the relationship existing between $\Velocity$ and $\CoeffTranstFToBGenerique$ by calculating the error made on the $Training~Database$ by the model parametrized with different $(\Velocity,\CoeffTranstFToBGenerique)$ pairs. To do so, we ranged the domain  $\Velocity \times \CoeffTranstFToBGenerique$ and we calculated the Relative Residual Sum of Squares ($RRSS$) (\ref{eq:RRSS}) associated with the models parametrized with different tested $(\Velocity,\CoeffTranstFToBGenerique)$ pairs:

\begin{gather}
\label{eq:RRSS}
\displaystyle{RRSS(\Velocity,\CoeffTranstFToBGenerique) = \sum\limits_{i=1}^n\left(\sum\limits_{j=1}^m\left(\frac{(y_{ij_{obs}}-y_{ij_{pred}}(\Velocity,\CoeffTranstFToBGenerique))}{y_{ij_{obs}}}\right)^2\right)},
\end{gather}

\noindent where $n$ corresponds to the number of individuals contained in the $Training~Database$ and $m$ the number of points on the curves. $y_{ij_{obs}}$ and $y_{ij_{pred}}$ correspond respectively to the observed and the predicted value of the  $j^{th}$ point of the $i^{th}$ individual. Therefore $RRSS$ corresponds to the sum of the squared relative differences between the predicted curves and the initially generated curves.\\

\begin{figure}[h!]
	\centering
\includegraphics[width=0.9\textwidth]{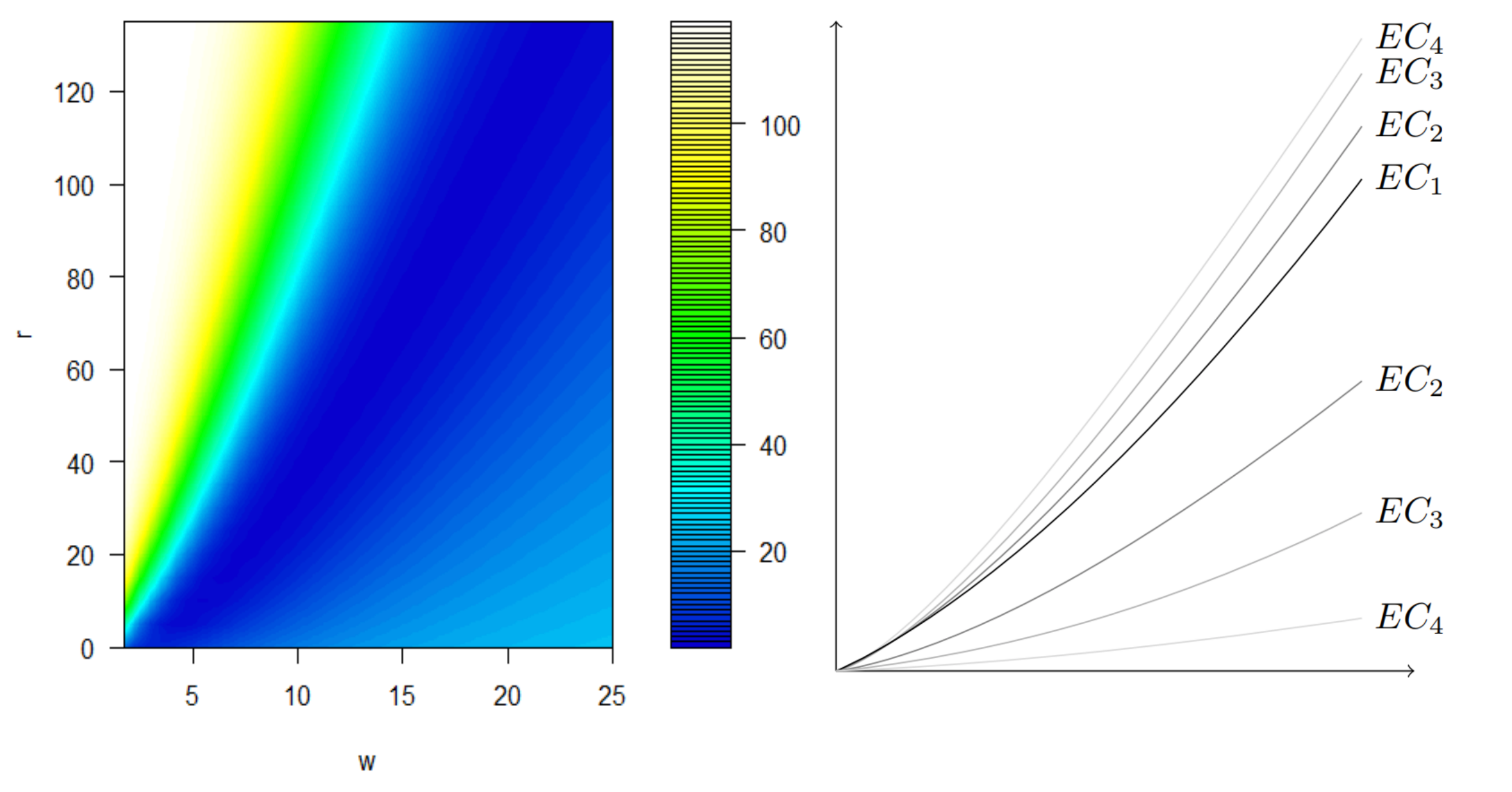}
	\caption{The value of the $RRSS$ according to $\Velocity$ and $\CoeffTranstFToBGenerique$ (a: left), and the schema of the different Equivalent Couples (EC) (b: right)}
		\label{fig:RRSSwr}
\end{figure}

\begin{figure}[h!]
	\centering
\includegraphics[width=0.8\textwidth]{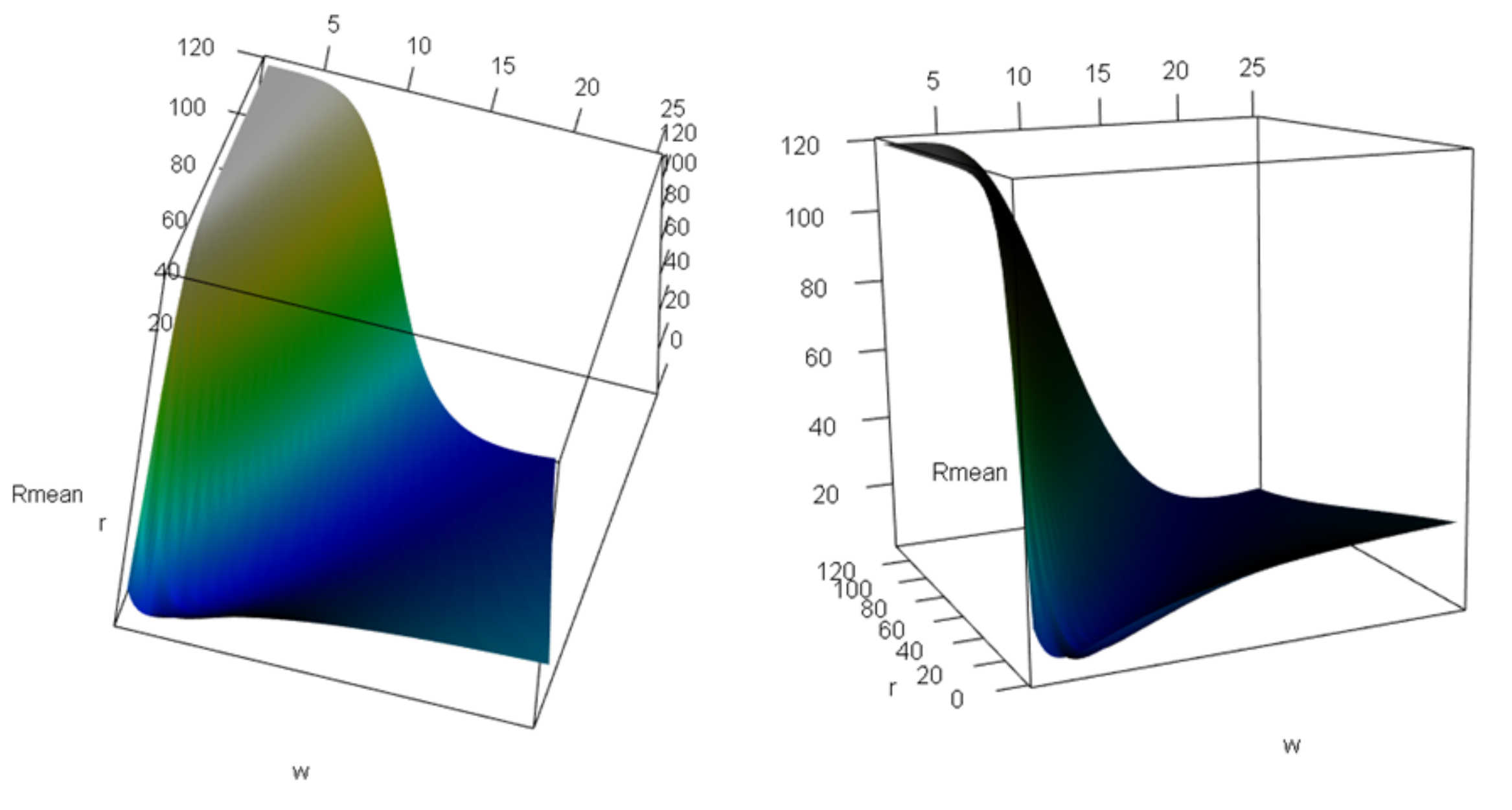}
	\caption{The $3D$ representation of the value of the $RRSS$ according to $\Velocity$ and $\CoeffTranstFToBGenerique$}
		\label{fig:RRSSwr3D}
\end{figure}

Figures \ref{fig:RRSSwr} and \ref{fig:RRSSwr3D} give the values of the RRSS according to the values of $\Velocity$ and $\CoeffTranstFToBGenerique$. The existence of a series of equivalent pairs - that is a series of pairs leading to the same value of $RRSS$ - can be seen in Figure \ref{fig:RRSSwr}(a). There is an area where the $RRSS$ values are lower (Figure \ref{fig:RRSSwr3D}), and corresponding to the curve $EC1$ in Figure \ref{fig:RRSSwr}(b). We assumed that the optimal ($\VelocityOpt$,$\CoeffTranstFToBGeneriqueOpt$) pair, inducing the lowest RRSS, belongs to this curve. Therefore, we set out to determine the equation of the curve $EC1$. \\

\subsection{Search for the $(\VelocityOpt,\CoeffTranstFToBGeneriqueOpt)$ pairs inducing the lowest $RRSS$}

To find the equation of the curve $EC1$ we sought for different values of $\Velocity$, the value of $\CoeffTranstFToBGenerique$
minimizing the $RRSS$ value. To do that, for each tested value of $\Velocity$ we used the optimization
algorithm DIRECT to find the value of $\CoeffTranstFToBGenerique$ minimizing the objective function,
\begin{gather}
\label{fObj_wr}
\displaystyle{f_{obj} (\CoeffTranstFToBGenerique)= \frac{1}{n}\sum\limits_{i=1}^n\left(\sum\limits_{j=1}^m\left(\frac{(y_{ij_{obs}}-y_{ij_{pred}}(\Velocity,\CoeffTranstFToBGenerique))}{y_{ij_{obs}}}\right)^2\right)},
\end{gather}

\noindent corresponding to the average RRSS.\par
\begin{figure}[h]
	\centering
\includegraphics[width=0.6\textwidth]{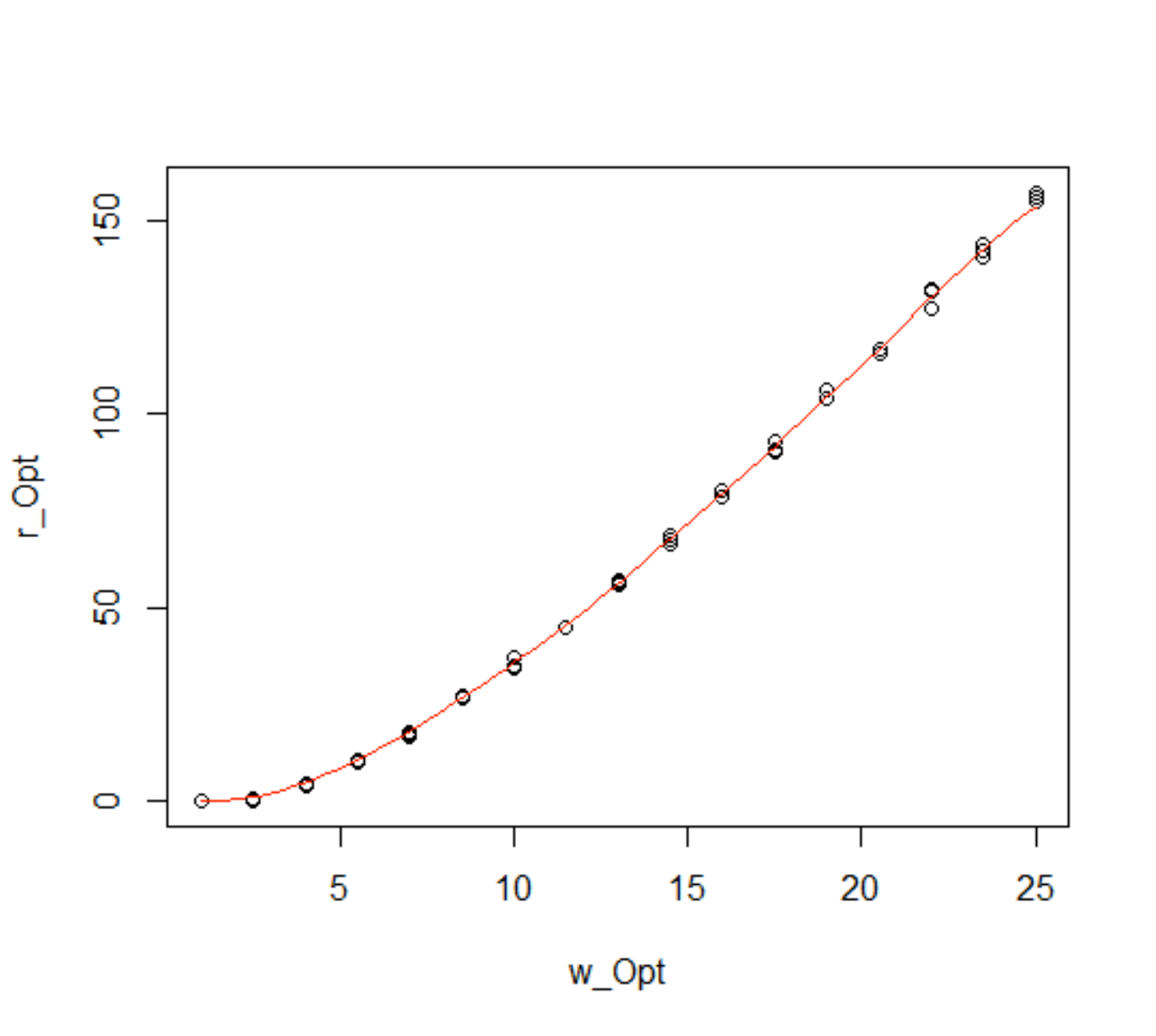}
	\caption{The Nadaraya-Watson kernel regression linking the ($\VelocityOpt$, $\CoeffTranstFToBGeneriqueOpt$) parameter pair.}
			\label{fig:RRSSmin}
\end{figure}

To obtain several fitted values of $\CoeffTranstFToBGenerique$ for each tested value of $\Velocity$, we sampled the $Training~Database$: we sampled $20$ curves from the $30$ test curves and we fitted $\CoeffTranstFToBGenerique$ on those $20$ selected curves. We ultimately obtained three values of $\CoeffTranstFToBGenerique$ for each tested value of $\Velocity$ (Figure \ref{fig:RRSSmin}). Using a Nadaraya-Watson kernel regression (See \cite{nadaraya1964estimating} and \cite{watson1964smooth}), we obtained a non-parametric relationship linking $\VelocityOpt$ and $\CoeffTranstFToBGeneriqueOpt$ in the form of:
\begin{gather}
\label{eq:model_r_w_App}
r_{opt}= \hat{m}(\omega_{opt}) + \epsilon,
\end{gather}

\noindent where $\hat{m}$ corresponds to the Nadaraya-Watson estimator.\par

Knowing the relationship between $\VelocityOpt$ and $\CoeffTranstFToBGeneriqueOpt$, it is possible to deduce one of these two parameters according to the value of the other parameter. Hence, this relationship reduces the number of parameters that need to be learned simultaneously.


\subsection{Study of the relationship between $\CoeffFixGenerique$ and $\CoeffUtilisation$}
\label{Subsection:Relation_f_u}

There is also a compensation effect between $\CoeffFixGenerique$ and $\CoeffUtilisation$: a high value of $\CoeffFixGenerique$ can be compensated by a low value of $\CoeffUtilisation$, and vice versa. \par
As above for $\Velocity$ and $\CoeffTranstFToBGenerique$, we sought to determine the relationship existing between $\CoeffFixGenerique$ and $\CoeffUtilisation$ to be able to deduce one of these two parameters according to the other one and further reduce the number of parameters to learn simultaneously.\\

\begin{figure}[h]
	\centering
\includegraphics[width=0.9\textwidth]{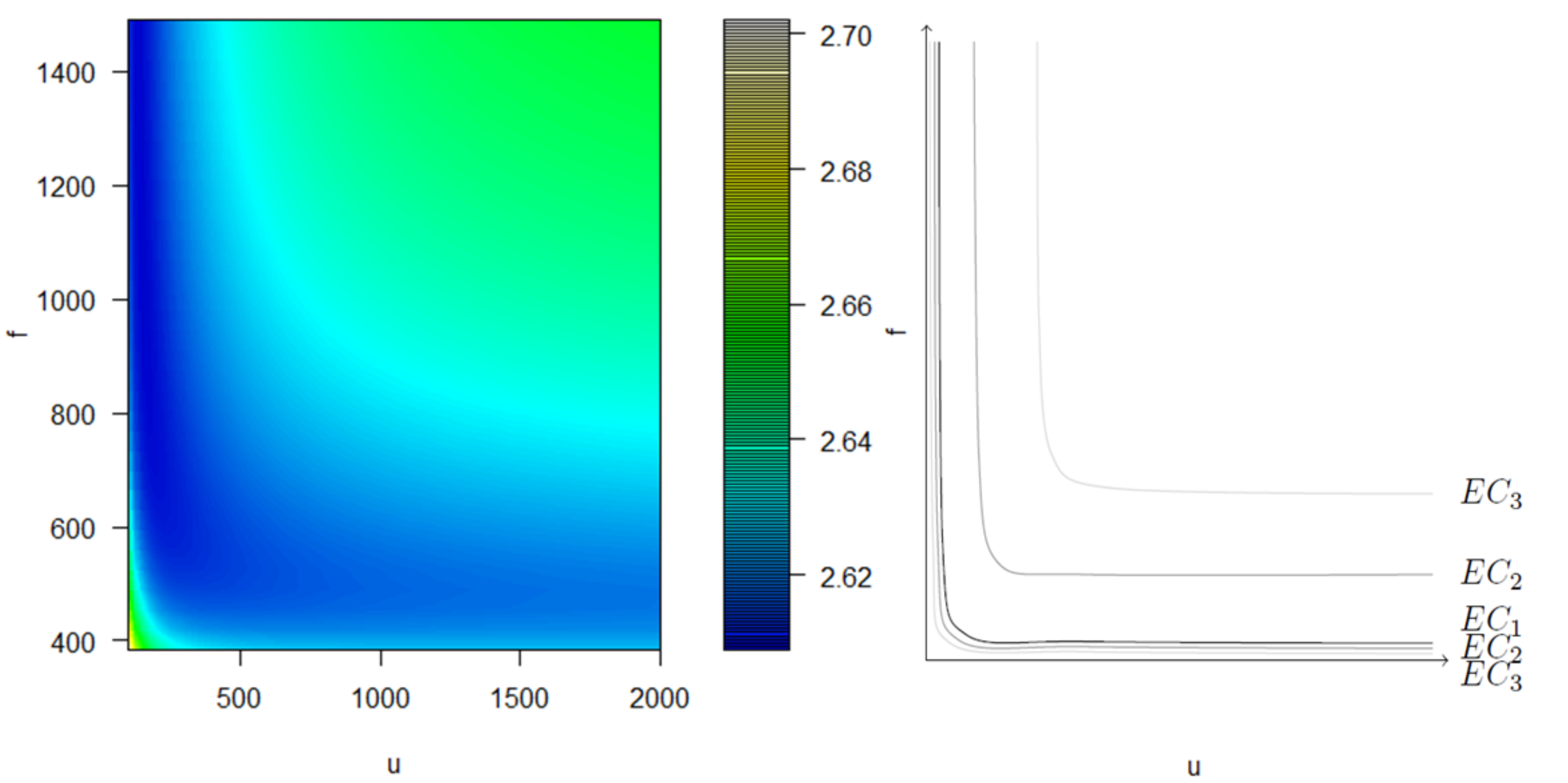}
	\caption{The value of the RRSS according to $\CoeffFixGenerique$ and $\CoeffUtilisation$ (a) and the schema of the different Equivalent Couples (EC) (b)}
		\label{fig:RRSSfu}
\end{figure}

\begin{figure}[h]
	\centering
\includegraphics[width=0.9\textwidth]{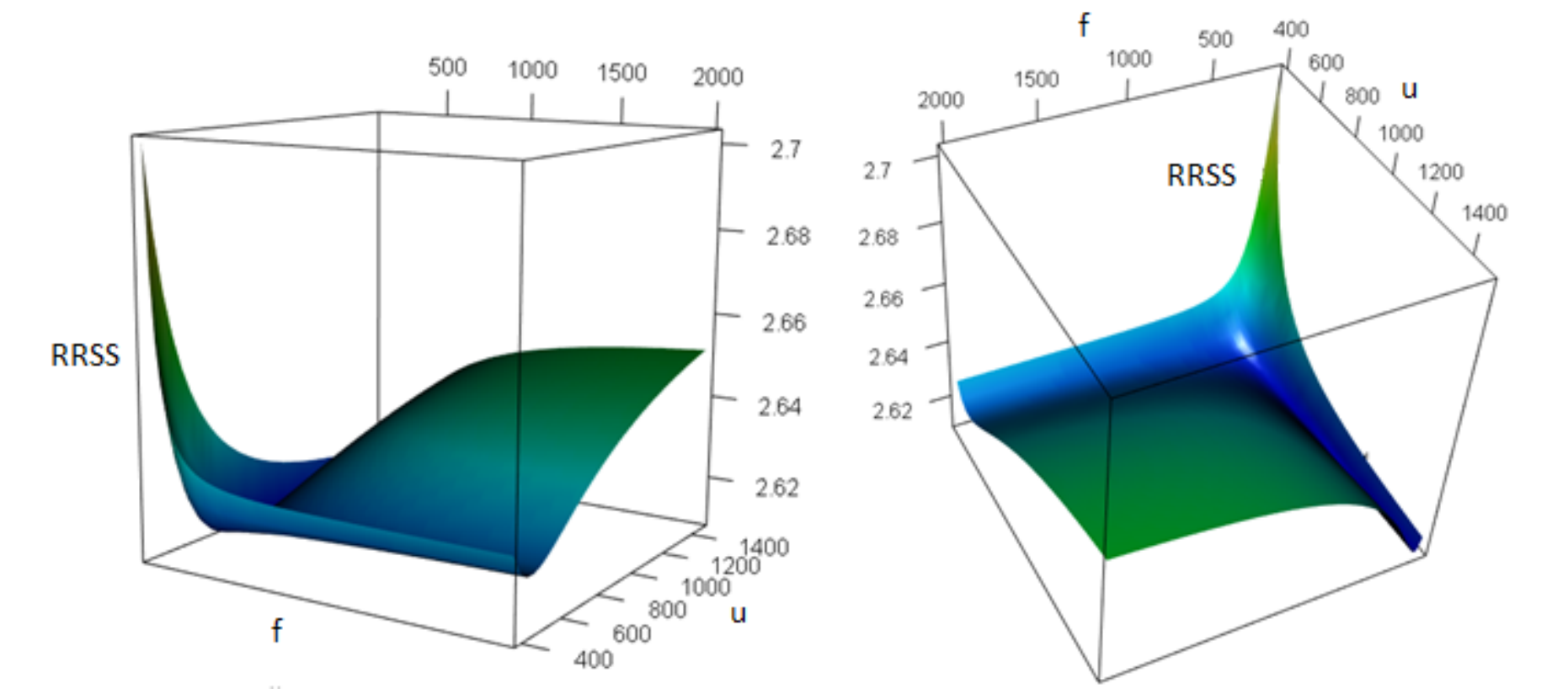}
	\caption{The $3D$ representation of the value of the RRSS according to $\CoeffFixGenerique$ and $\CoeffUtilisation$}
			\label{fig:RRSSfu3D}
\end{figure}

As above, we ranged the domain $\CoeffFixGenerique \times \CoeffUtilisation$ and calculates the $RRSS$ of the models parameterized with different pairs of values for $(\CoeffFixGenerique, \CoeffUtilisation)$ (Figures \ref{fig:RRSSfu} and \ref{fig:RRSSfu3D}). This study demonstrates a series of equivalent pairs. There is an area where the $RRSS$ values are lower (Figure \ref{fig:RRSSfu3D}) and corresponding to the $EC1$ curve in Figure \ref{fig:RRSSfu}(a). We assumed that the optimal $(\CoeffFixGeneriqueOpt, \CoeffUtilisationOpt)$ pair inducing the lowest $RRSS$, belongs to this curve. Therefore, we set out to determine the equation of this curve.

\subsection{Search for the $(\CoeffFixGeneriqueOpt, \CoeffUtilisationOpt)$ pairs that lead to the lowest $RRSS$}

To find the equation of the curve $EC1$ associated with the lowest $RRSS$, we looked for the value of  $\CoeffUtilisation$ minimizing the $RRSS$ value for different given values of $\CoeffFixGenerique$. For each value of $\CoeffFixGenerique$, we used the optimization algorithm DIRECT to find the value of  $\CoeffUtilisation$ minimizing the objective function (\ref{fObj_fu}) corresponding to the average $RRSS$.
\begin{gather}
\label{fObj_fu}
\displaystyle{f_{obj} (\CoeffUtilisation)= \frac{1}{n}\sum\limits_{i=1}^n\left(\sum\limits_{j=1}^m\left(\frac{(y_{ij_{obs}}-y_{ij_{pred}}(\CoeffFixGenerique,\CoeffUtilisation))}{y_{ij_{obs}}}\right)^2\right)}.
\end{gather}

\begin{figure}[th]
	\centering
\includegraphics[width=0.7\textwidth]{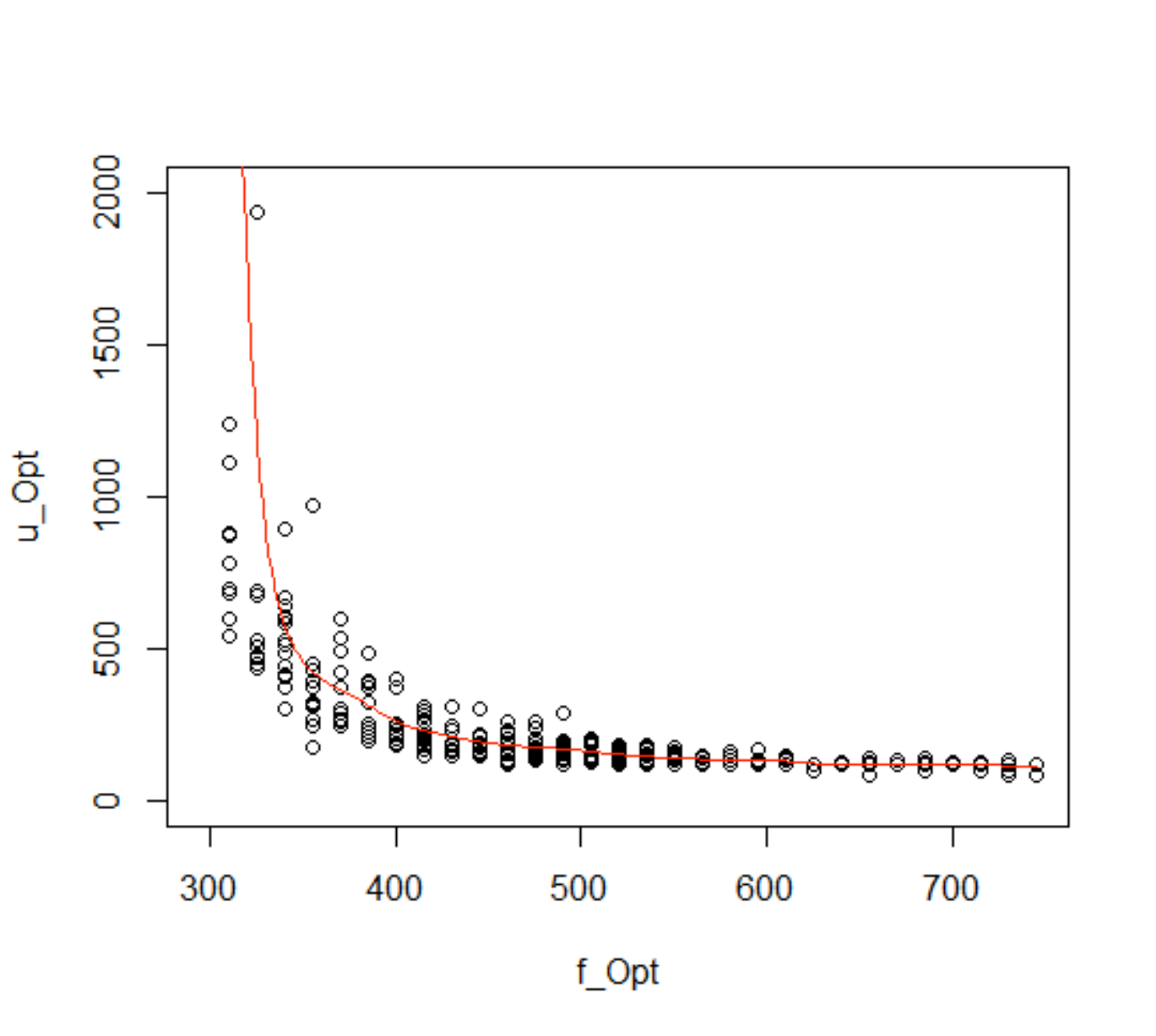}
	\caption{The Nadaraya-Watson kernel regression linking $\CoeffFixGeneriqueOpt$ and $\CoeffUtilisationOpt$.}
	\label{fig:RRSSmin_fu}
\end{figure}

As above, to obtain several fitted values of $\CoeffUtilisation$ for each tested value of $\CoeffFixGenerique$, we sampled the $Training~Database$. At the end of the fitting, we obtained three values of $\CoeffUtilisation$ for each tested value of $\CoeffFixGenerique$ (Figure  \ref{fig:RRSSmin_fu}). Using a Nadaraya-Watson kernel regression, we obtained a non-parametric relationship linking $\CoeffFixGeneriqueOpt$ and $\CoeffUtilisationOpt$ in the form of:
\begin{gather}
\label{eq:model_r_w_App}
\CoeffUtilisationOpt= \hat{m}(\CoeffFixGeneriqueOpt) + \epsilon,
\end{gather}

\noindent where $\hat{m}$ corresponds to the Nadaraya-Watson estimator.\\

Knowing the relationship existing between $\CoeffFixGeneriqueOpt$ and $\CoeffUtilisationOpt$, it is possible to deduce one of these two parameters according to the value of the other parameter. Hence, this relationship further reduces the number of parameters to learn simultaneously.

\end{document}